\documentclass[9pt,twocolumn]{article}
\usepackage{hyperref}
\usepackage{amsmath}
\newenvironment{alphafootnotes}
  {\par\edef\savedfootnotenumber{\number\value{footnote}}
   
   \setcounter{footnote}{1}}
  {\par\setcounter{footnote}{\savedfootnotenumber}}

\usepackage{fullpage}

\usepackage{tikz}
\usetikzlibrary{arrows, decorations.markings,
  decorations.pathmorphing, shapes, fit, backgrounds, patterns,
  calc,petri,snakes}

\tikzstyle{decay} = [draw, ->, decorate, decoration = {snake,
  amplitude = 0.4mm, segment length=2mm, post length = 2mm, pre length
= 1mm}]

\tikzstyle{default} = [draw, minimum size = 3em, text width = 4em, text centered]
\tikzstyle{wide}=[draw, minimum size=3em, text width=7.5em, text
centered]
\tikzstyle{shortbox}=[draw, minimum size=2em, text width=4.5em, text centered]
\tikzstyle{bigbox}=[draw, inner sep=20pt,label={[align=right,shift={(-1.5ex,3ex)}]south east:\llap{#1}}]
\tikzstyle{box} = [draw, minimum size=2em, text width=1.5em, text centered]
\definecolor{colorS}{RGB}{0,154,128}
\definecolor{colorI}{RGB}{255,32,0}
\definecolor{colorR}{RGB}{205,10,179}

\usepackage{todonotes}

\renewcommand{\vec}[1]{\mathbf{#1}}
\newcommand{\ave}[1]{\left\langle #1 \right \rangle}
\newcommand{\order}{\mathcal{O}}
\newcommand{\erdosrenyi}{Erd\H{o}s--R\'{e}nyi}

\newcommand{\pd}[2]{\frac{\partial #1}{\partial #2}}
\newcommand{\pds}[3][2]{\frac{\partial^{#1} #2}{\partial #3^{#1}}}

\newcommand{\Lap}{\mathcal{L}}
\newcommand{\Ro}{\mathcal{R}_0}

\newcounter{quotecount}
\renewcommand{\thequotecount}{\roman{quotecount}}
\newcommand{\MyQuote}[1]{%
\refstepcounter{quotecount}%
\indent\parbox{\columnwidth - 4\parindent}{\bigskip\em #1}\hfill(\thequotecount)\bigskip}

\title{
Random Spatial Networks: Small Worlds without Clustering, Traveling Waves, and Hop-and-Spread Disease Dynamics
}

\author{John Lang, Hans De Sterck, Jamieson L. Kaiser, Joel C. Miller}

\begin{document}
%%%%%%%%%%%%%%%%%%%%%%%%%%%%%%%%%%%%%%%%%%%%%%
%%%%%%%%%%%%%%%%%%%%%%%%%%%%%%%%%%%%%%%%%%%%%%
%%%%%%%%%%%%%%%%%%%%%%%%%%%%%%%%%%%%%%%%%%%%%%
\twocolumn[
  \begin{@twocolumnfalse}
    \maketitle
    \begin{abstract}
%Please provide an abstract of no more than 250 words in a single paragraph. Abstracts should explain to the general reader the major contributions of the article. References in the abstract must be cited in full within the abstract itself and cited in the text.\\
Random network models play a prominent role in modeling, analyzing and understanding complex phenomena on real-life networks. However, a key property of networks is often neglected: many real-world networks exhibit spatial structure, the tendency of a node to select neighbors with a probability depending on physical distance.
Here, we introduce a class of random spatial networks (RSNs) which generalizes many existing random network models but adds spatial structure. In these networks, nodes are placed randomly in space and joined in edges with a probability depending on their distance and their individual expected degrees, in a manner that crucially remains analytically tractable.
We use this network class to propose a new generalization of small-world networks, where the average shortest path lengths in the graph are small, as in classical Watts-Strogatz small-world networks, but with close spatial proximity of nodes that are neighbors in the network playing the role of large clustering.  Small-world effects are demonstrated on these spatial small-world networks without clustering.
We are able to derive partial integro-differential equations governing susceptible-infectious-recovered disease spreading through an RSN, and we demonstrate the existence of traveling wave solutions. If the distance kernel governing edge placement decays slower than exponential, the population-scale dynamics are dominated by long-range hops followed by local spread of traveling waves. This provides a theoretical modeling framework for recent observations of how epidemics like Ebola evolve in modern connected societies, with long-range connections seeding new focal points from which the epidemic locally spreads in a wavelike manner.
   \end{abstract}
  \end{@twocolumnfalse}
]
%%%%%%%%%%%%%%%%%%%%%%%%%%%%%%%%%%%%%%%%%%%%%%
\section{Introduction}
%%%%%%%%%%%%%%%%%%%%%%%%%%%%%%%%%%%%%%%%%%%%%%

The spread of infectious disease through human and animal populations exhibits a range of patterns.  In the pre-vaccination era, Measles in the UK spread out from London in a coherent spatial pattern~\cite{grenfell2001travelling}.  Other diseases exhibiting such wavelike behavior include rabies in racoons~\cite{childs2000predicting,biek2007high} and vampire bats~\cite{benavides2016spatial} as well as the Black Death of 1347-1351 in Europe which claimed an estimated 30-50\% of the European population~\cite{bos2011draft,christakos2007recent}.

\begin{alphafootnotes}
These coherent spatial patterns seem the norm in many epizootics as well as historical human epidemics.  However, more recent human epidemics have exhibited different patterns.  In modern connected societies there are long-range connections facilitated by travel infrastructure that play increasingly important roles in disease propagation \cite{balcan2009multiscale}.  These lead to the appearance of new spatially dissociated locally spreading clusters of disease.  We will refer to this pattern of spread as ``hop-and-spread'' dynamics. The 2013-2016 epidemic of Ebola virus disease in West Africa had significant long-range hops: sequencing of over 1600 Ebola virus genomes reveals a heterogeneous and spatially dissociated collection of transmission clusters of varying size, duration and 
%connectivity~\cite{dudas2016virus}.\footnote{Since \cite{dudas2016virus} has not been published yet, we include this preliminary reference to a visualization of some of the main findings of \cite{dudas2016virus}: \url{https://vimeo.com/152494592}.}
connectivity~\cite{dudas2016virus}.\footnote{ \url{https://vimeo.com/152494592} provides  a visualization of the processes described in \cite{dudas2016virus}.}
SARS and pandemic 2009 H1N1 influenza showed significant local spread, but the global dynamics were dominated by sporadic long-range transmission events.
\end{alphafootnotes}

Network models are a valuable tool for theoretical investigation of how the contact structure of a population governs the spread of infectious disease~\cite{newman:spread,meyers:contact,pastor2014epidemic}.  They also appear in many other contexts, such as understanding activation patterns in neurons~\cite{bullmore2009complex,o2013spreading}.  These networks also have spatial structure.  In particular, recent work on cortical networks has shown that the macaque cortex has strong structural specificity in terms of the strength of connections as a function of the distance between areas of the cortex \cite{ercsey2013predictive}. A simple spatial model of cortex connectivity predicts the existence of a strong core-periphery organization. Applying the spatial model to different animals leads to the suggestion that the mammalian cortex exhibits universal spatial architectural principles \cite{horvat2016spatial}. %This is another clear example of a complex network where spatial structure is important.

The behavior that emerges as a dynamic process spreads in a network comes from a combination of the process-specific rules governing the node-node interactions and the structure of the network which provides the underlying substrate along which the process spreads.  Often the structure of the network both in terms of connectivity structure and spatial structure plays a dominant role. Thus similar large-scale dynamics are observed for processes with different interaction rules as long as the underlying networks are similar.

\subsection{Real-World Spatial Networks}
The networks of direct human-human contacts and neuronal contacts mentioned above exhibit preferential connections between nearby nodes.  Spatial structure appears in many other network contexts as well~\cite{barthelemy2011spatial}.  These include human communications across mobile networks~\cite{lambiotte2008geographical}, wireless sensor networks~\cite{haenggi2009stochastic}, protected plant/animal habitats~\cite{hanski2000metapopulation}, wildlife interaction networks~\cite{davis2015spatial,hamede2009contact}, and even the physical internet~\cite{yook2002modeling}.  All of these networks demonstrate that shorter-range connections are preferred.

There is relatively little theoretical study of how spatial structure in a network affects  spreading processes.  This is largely because the available classes of spatial network models have a number of weaknesses.  In particular, they are not amenable to analytic investigation.  By way of contrast, non-spatial networks such as Chung-Lu networks~\cite{chung:connected} and Configuration-Model networks~\cite{newman:structurereview} provide significant insight.  This is largely because they have a ``locally tree-like'' structure, that is, for fixed $D$, the probability that a random node is in a cycle of length $D$ or less goes to zero as the number of nodes goes to infinity.  

The dynamics of many spreading processes, for example, complex contagions~\cite{centola:cascade}, the generalized epidemic process~\cite{janssen:GEP} and the Watts Threshold Model~\cite{watts:WTM} can be studied exactly in locally tree-like networks~\cite{miller2015complex,miller:contagion}.  There is a need for a similarly tractable model of spatial networks to allow us to study how spatial structure affects spreading dynamics.

\subsection{A Random Spatial Network Model}
To address this need we introduce a class of random spatial networks (RSNs) modeled after Chung-Lu networks.  Each node is assigned an expected degree, and edges are placed between nodes with probability proportional to the product of their expected degrees and a distance kernel.

RSNs are a subclass of inhomogeneous random graphs~\cite{soderberg2002general,bollobas2007phase} and generalize many well-known network models, including Random Geometric Graphs, Chung--Lu networks, and Newman--Watts networks.  

In particular, we will focus on the relation between RSNs and small-world networks. Small-world networks are considered to be highly-clustered networks (i.e., they have many short cycles), but the typical path lengths between randomly chosen node pairs are small.  We will see that, in certain limits, RSNs exhibit these properties, and processes spreading in these RSNs mimic those of well-known small-world network models.  

By increasing the node denisty, we can tune RSNs to have the same geometric distances between neighbors (the same ``spatial structure''), but small clustering.  In the unclustered limit, we still see many of the same behaviors, suggesting that ``small-world'' behaviors on real-world networks with spatial structure may be consequences of the distribution of long-distance and short-distance links rather than consequences of the existence of long-distance and clustered links.  By varying the clustering while preserving the spatial structure, we can disentangle which properties of spreading processes in small-world networks are consequences of clustering and which are explained by spatial proximity.

As a specific application, we will study RSNs to explore hop-and-spread dynamics of susceptible-infectious-recovered (SIR) disease spread.  We will see this behavior in networks that satisfy the small-world property of high clustering with short network diameters, but the same behavior emerges in RSNs with short network diameters and negligible clustering, but with high spatial proximity of neighbors in the graph.  

As the node density increases in RSNs, stochastic effects in disease propagation models become less important, and we are able to derive differential equations that govern the spatial dynamics of SIR disease on RSNs. Using these equations, we are able to demonstrate the existence of nonlinear traveling wave solutions that retain their shapes. We derive the wave speed of the traveling waves and show that no finite wave speed exists if the distance kernel governing the probability of connections existing at different lengths decays slower than exponentially. We demonstrate that the traveling wave solutions in the numerical differential equation models closely match the traveling wave structures arising in stochastic simulations of SIR disease on the RSNs. 

A major advantage of RSNs compared to other networks with spatial structure is the suitability of the networks to analytic results in the high-density limit .  Although we demonstrate this only for disease spread, many analytic techniques used to study other spreading processes in random networks will also apply to RSNs.

%Other processes, such as the spread of a complex contagion which requires transmissions from multiple neighbors are likely to be more dependent on the network's clustering~\cite{centola:weakness}.

%%%%%%%%%%%%%%%%%%%%%%%%%%%%%%%%%%%%%%%%%%%%%%
\section{The Random Spatial Networks Class}
%%%%%%%%%%%%%%%%%%%%%%%%%%%%%%%%%%%%%%%%%%%%%%

\begin{figure}
\includegraphics[width=\linewidth]{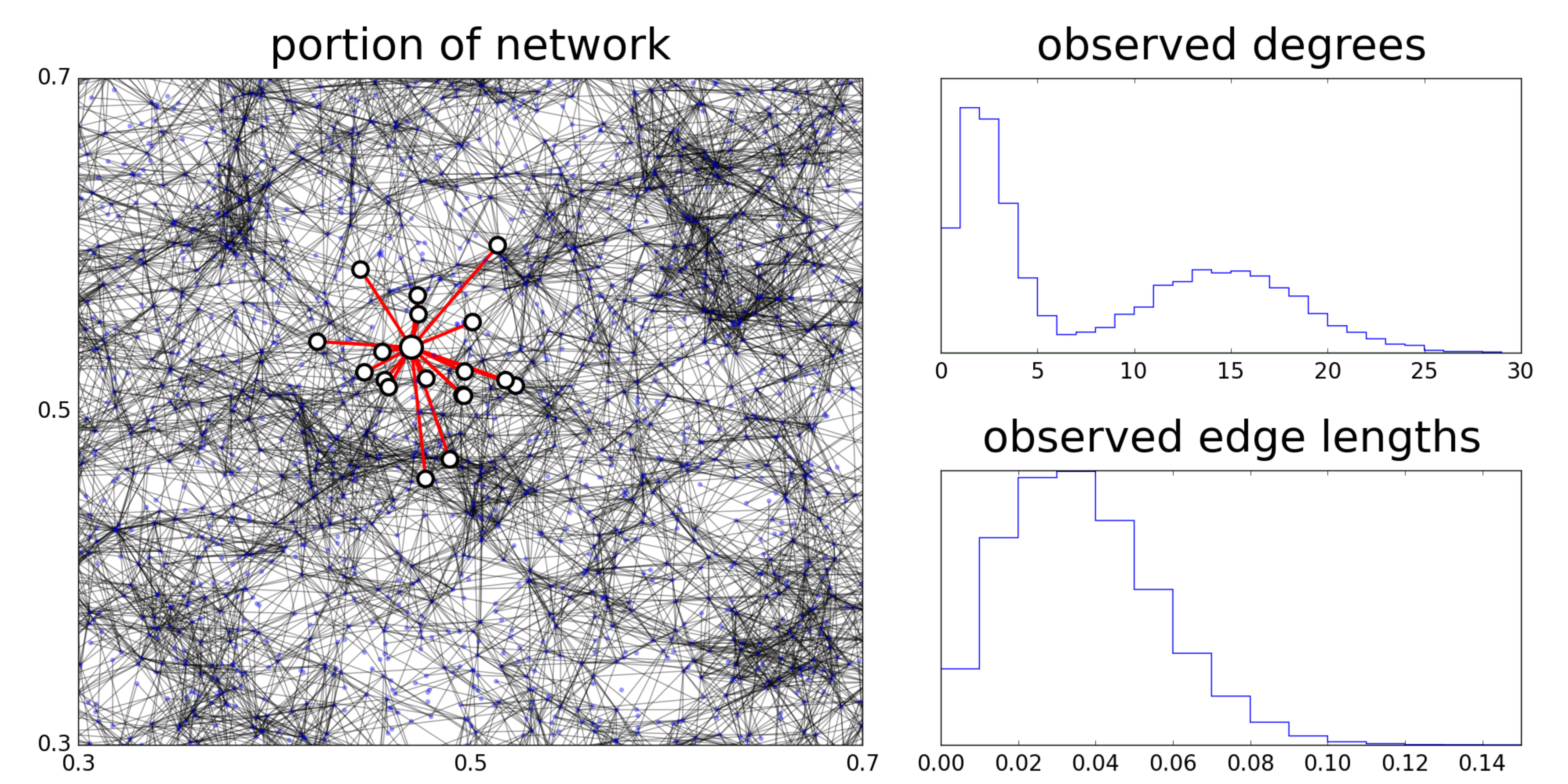}\\
\caption{\small An example RSN and its properties. The imposed degree distribution is bimodal, and the distance kernel is Gaussian, implying that all network connections will be local. One node and its neighbours are highlighted. A random network without spatial structure would show neighbours throughout the domain.}
\label{fig:example}
\end{figure}

%It has been clearly demonstrated that heterogeneity in degree plays a large role in many spreading processes on networks~\cite{XXX}.  It is also clear that the small-world structure of a population can also play an important role~\cite{XXX}.  The classical small-world networks do not include heterogeneous degree.  By a small adaptation of the random spatial networks we have introduced here, we can produce heterogeneous networks with controllable spatial structure and clustering.

%To create these networks, we take some space $V$ and assign nodes to random locations, with density $\rho$ nodes per unit volume of $V$.  The node $u$ is given an expected degree $\kappa_u$ from a probability distribution.  An edge is placed between nodes $u$ and $v$ with probability $\kappa_u \kappa_v f(d_{uv}) / \rho\ave{\kappa}$ for some function $f$.  Each edge exists independently of any other.  In the limit $\rho \to \infty$, the degree of a node with a given $\kappa$ will be chosen from a Poisson distribution with mean $\kappa$.\footnote{We can also use a regular grid layout for the nodes, but in this case the observed degree of a node may not be from a Poisson distribution.}

We now define our proposed random spatial network class.  
%We distribute nodes in space randomly and generate edges between pairs of nodes with probabilities that are proportional to the desired degrees of the nodes (as in the Chung--Lu model), but also depend on the distance between the nodes.  
To generate a random spatial network, we first randomly place nodes into some region $V$ of a Euclidean space with some density $\rho$, which for our purposes is $\rho = N/|V|$, where $N$ is the total number of nodes and $|V|$ is the (assumed finite) volume of the space $V$.

Each node $u$ is independently assigned an expected degree $\kappa_u$ from some distribution $P(\kappa)$.  
%We use $P(\kappa)$ as a shorthand for $P(\kappa_u=\kappa)$, the probability $\kappa_u=\kappa$.  
The average of $\kappa$ is denoted $\ave{\kappa}$.  We assume that the \emph{distance kernel} $f$ is non-negative and integrates to $1$: $\int_V f(|\vec{x}|) \, \mathrm{d}\vec{x} = 1$.  Typically we also assume that $f$ decreases monotonically.

An edge is placed between nodes $u$ and $v$ with probability 
%------------------------------------
\begin{equation}
p_{uv} =  \min\left(\kappa_u\kappa_v \frac{f(d_{uv})}{\rho \ave{\kappa}}, 1\right),
\label{eq:p}
\end{equation}
%------------------------------------
If $p_{uv}<1$ always and boundary effects in $V$ are negligible then the expected degree of a node $u$ is
\[
\int_V \int_0^\infty \rho f(|\vec{x}_v-\vec{x}_u|) \kappa_u \kappa_v \frac{P(\kappa_v)}{\rho \ave{\kappa}} \mathrm{d}\kappa_v \, \mathrm{d}\vec{x}_v = \kappa_u \ .
\]
In the large $\rho$ limit, the observed degrees of nodes with a given expected degree $\kappa$ is Poisson-distributed with mean $\kappa$.  

We will generally take $V$ to be the unit interval $[0,1]$, a ring $[0,1]$ with the end points set equal, the square $[0,1]\times [0,1]$, or the torus $[0,1]\times [0,1]$ with periodic boundaries. 

This model offers considerable flexibility:
\begin{itemize}
\item The distance kernel can be tuned to generate a wide range of spatial structure. 
\item The choice of the distribution of expected degrees allows us to tune the distribution of realized degrees.
\item Many further generalizations (presented in the discussion in Section \ref{sec:conclusion}) emerge naturally.
\end{itemize}
Figure~\ref{fig:example} shows part of an RSN with $10^4$ nodes in $[0,1]\times[0,1]$ with an imposed bimodal degree distribution and a Gaussian distance kernel.

RSNs contain several widely-studied models as special cases and are themselves a type of inhomogeneous random graph~\cite{soderberg2002general,bollobas2007phase}.  They incorporate both degree heterogeneity and spatial structure
in a way that, crucially, remains analytically tractable.
%, including Random Geometric Graphs~\cite{penrose2003random}, Chung--Lu random graphs~\cite{chung:connected}, Small world networks~\cite{watts:collective,newman1999renormalization}, Long-range percolation~\cite{X}, and Waxman graph~\cite{X}.

%or into a lattice with density $\rho$.
%We concentrate on the assumption that nodes are randomly placed in $V$.

%\JCM{I put a page break here because there is a crazy bug going on that wouldn't go away without it.  I think it's because of the figure coming close to a line break}

%The RSN model is notable because it has a flexible degree distribution (as do many existing models), but it adds the essential ability to generate connections depending on spatial proximity in a versatile way, and it does so in a way that it remains amenable to analytic techniques. It is the first model to combine these key properties. 
Before applying RSNs to model SIR disease spread, we first discuss some of their general properties, including how they relate to existing random graph models. 

%------------------------
\subsection{Relation to Existing Random Network Models}
%------------------------
The \erdosrenyi{} network class (actually two subtly different classes one introduced by Gilbert~\cite{gilbert1959random} and another by Erd\H{o}s and R\'{e}nyi~\cite{erdos1959random}) is the oldest and most famous random network model.  Its degree distribution is homogeneous, and it does not have spatial structure.  The more recent Molloy--Reed~\cite{molloy1995critical} and Chung--Lu~\cite{chung2002average} network classes incorporate a heterogenous degree distribution. These two models are closely-related to \erdosrenyi{} networks and also do not have spatial structure.  The ``locally tree-like'' structure of these network classes permits a range of analytic techniques.
%\HDS{Joel, can you give some insight into what ``locally tree-like'' means, in one sentence? Is this related to the edges being present independent of one another?}

The Exponential Random Graph model can handle degree heterogeneity~\cite{robins2007introduction} and some spatial structure~\cite{wong2006spatial}, but typically the actual networks considered are  small because they are expensive to generate.  Also there are no known analytic methods to study spreading processes.

Some other network models also incorporate a degree of spatial structure.  Among these are Waxman networks~\cite{waxman1988routing}, Spatially-Embedded Random Networks~\cite{barnett2007spatially},  Random Geometric Graphs~\cite{penrose2003random}, Long-range Percolation~\cite{newman1986one}, and various  ``spatially-constrained networks''~\cite{kosmidis2008structural,vladimirov2011wave}. Often these are used to understand patterns emerging in the brain~\cite{o2013spreading} or wireless sensor networks~\cite{haenggi2009stochastic}. 
Even though this is normally not emphasized, the prototypical small-world networks of \cite{watts1998collective,newman1999renormalization} also implicitly feature spatial structure, as is apparent in the ring-shaped graphical way these networks are normally represented. For these networks, this spatial structure is entangled with the clustering of the network. 
These existing spatial models have significant weaknesses: 
\begin{itemize}
\item None of these models incorporate degree heterogeneity.  
\item  Many place nodes on lattice points and assign edges based on distance.  The number of edges in grid-aligned or diagonal directions may differ significantly, leading to (often unrecognised) anisotropic spread.
\item Many of these models have difficult-to-control correlations because of significant clustering.  The microscopic structure makes analytic investigation of dynamic processes difficult.  In particular for small-world networks we cannot separate the effects of spatial structure from clustering.
\end{itemize}
Due to this last weakness, almost all studies of spreading processes in networks with spatial structure are limited to simulation. When equation-based analysis is attempted, it lacks quantitative agreement with simulations~\cite{vladimirov2011wave}.

A recent model with similarity to ours is the Geometric Inhomogeneous Random Graph model~\cite{bringmann2016geometric}.  This model includes spatial structure and degree heterogeneity, but the analysis is limited to distance kernels that are a power of the distance.

In Section \ref{sec:special-cases} of the Supporting Information (SI) we show that many existing random graph models arise as special cases of our RSN class. For the Newman--Watts small-world networks \cite{watts:collective,newman1999renormalization} we give the details of this equivalence here, since the ability of the RSN framework to produce classical (clustered) small-worlds networks will feature in the next section where we introduce a new class of small-world networks without clustering.

%------------------------
\subsection{Newman--Watts Small-World Networks as RSNs}
%------------------------
We can exactly recover the Newman--Watts small-world network model \cite{watts:collective,newman1999renormalization} in the RSN framework if we place the nodes at uniform distances.  In the Newman-Watts small-world network model, $N$ nodes are arranged in a ring, and each is connected to the nearest $k/2$ neighbors on either side where $k \ll N$ (resulting in a degree of $k$).  Then each pair of nodes which is not already in an edge is joined together with some small probability $p$ so that the average degree is $k+\epsilon$ where $\epsilon = p(N-k-1)$.

To recover this as an RSN, we place $N$ nodes exactly at uniform intervals in the one-dimensional ring $[0,1]$ with periodic boundaries.  So $\rho = N$ and distances are integer multiples of $1/N$.  We assign each node an expected degree $\kappa = k+\epsilon$. 
We set our distance kernel $f$ to be $f(d_{uv})=N/(k+\epsilon)$ for $d_{uv}\leq k/(2N)$ which results in each node being connected to its nearest $k/2$ neighbors on each side.  Then we set $f(d_{uv}) = Np/(k+\epsilon)$ for $d_{uv}>k/(2N)$. This results in all other node pairs being connected with probability $p$.
These probabilities are exactly the same as in the Newman--Watts model.

%%%%%%%%%%%%%%%%%%%%%%%%%%%%%%%%%%%%%%%%%%%%%%%%%%%%%%
\section{RSNs and Small-World Networks}
%%%%%%%%%%%%%%%%%%%%%%%%%%%%%%%%%%%%%%%%%%%%%%%%%%%%%%
%\JCM{if move 2.B into supplement, mention here that the Newman--Watts networks can be exactly reproduced by choice of distance kernel and placing nodes at regular intervals.}
In the traditional small-world networks of Watts-Strogatz~\cite{watts:collective} and Newman-Watts~\cite{newman1999renormalization} the networks are designed to be highly clustered and to contain long-range connections which result in short paths between seemingly far separated nodes.  Small-world networks are defined as having both a typical path length between nodes comparable to a random network and clustering much higher than a random network.  This is highlighted in the original publication as:
\MyQuote{We find that these systems can be highly clustered, like regular lattices, yet have small characteristic path lengths, like random graphs. We call them small-world networks.~\cite{watts:collective}
\label{quote:sw}}
In~\cite{watts1999networks} ``small-world networks'' were introduced as networks for which\\
\MyQuote{almost every
element of the network is somehow “close” to almost every other element,
even those that are perceived as likely to be far away.~\cite{watts1999networks}
\label{quote:close}}

\noindent
and it was argued that networks satisfying this must be highly clustered.

We propose a new class of networks that arise from the RSN model and satisfy the small-world property in quote~(\ref{quote:close}), but have negligible clustering.  We will refer to these as ``unclustered small-world networks'' and use the term ``classical small-world networks'' for networks which satisfy the classical, clustered, definition.  We will show that for some dynamical processes, the same behavior occurs in both unclustered and classical small-world networks.  

Classical small-world networks have been widely studied, often with an emphasis on the impact of the structure on spreading processes, such as disease~\cite{moore2000epidemics} or rumors~\cite{zanette2002dynamics} in a social network or firing activity in an epileptic seizure~\cite{netoff2004epilepsy,ponten2007small}.

However, the usual prototypical models based on taking a ring or a lattice and adding long-range connections have an implicit geometry, which produces a spatial structure.  In such networks, most nodes that are nearby in a network sense are also physically close.  This leads to the question of whether the behavior of some spreading process on a classical small-world network is a consequence of the clustering or of the spatial proximity of neighbors.  The roles of clustering and spatial proximity cannot easily be disentangled with standard small-world models of~\cite{newman1999renormalization,watts:collective}, even though we will see that these are distinct properties.

We will use RSNs to separate the impacts of clustering and spatial proximity.  By choosing $N_0$ and a distance kernel appropriately, we can generate an RSN with $N=N_0$ nodes that satisfies the standard definition of a small-world network (with clustering).  However, as we increase $N$ but keep the same distance kernel, the clustering coefficient scales like $1/N$ and becomes negligible at large $N$, but the spatial separation properties of neighboring nodes remain the same.  This allows us to interpolate between unclustered and classical small-world networks by simply increasing the density of nodes while holding all other properties the same.  In this unclustered limit, the networks become locally tree-like and many analytic tools become available.

\subsection{Classical Small-World RSNs}

%------------------------
\begin{figure*}[t]
\begin{center}
\hspace*{\fill}\includegraphics[width=0.7\columnwidth]{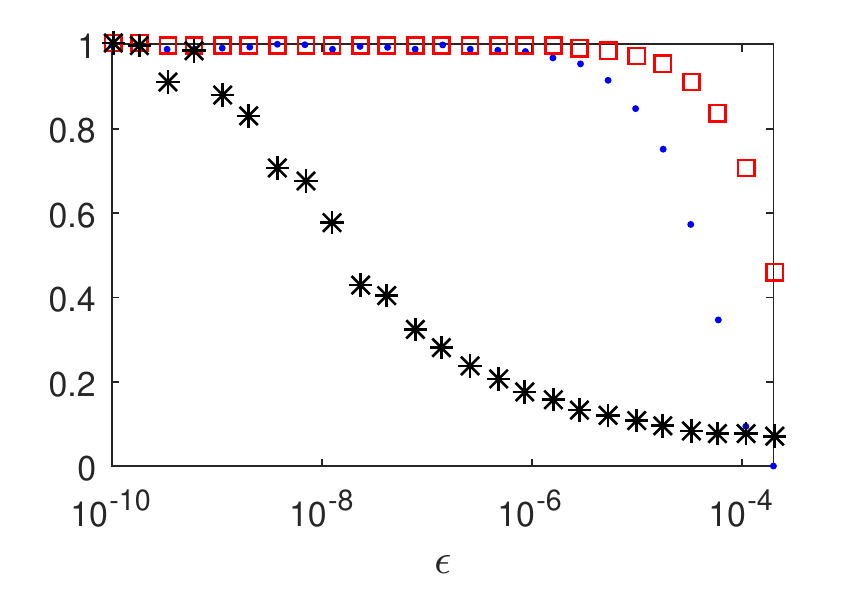}\hfill
\includegraphics[width=0.7\columnwidth]{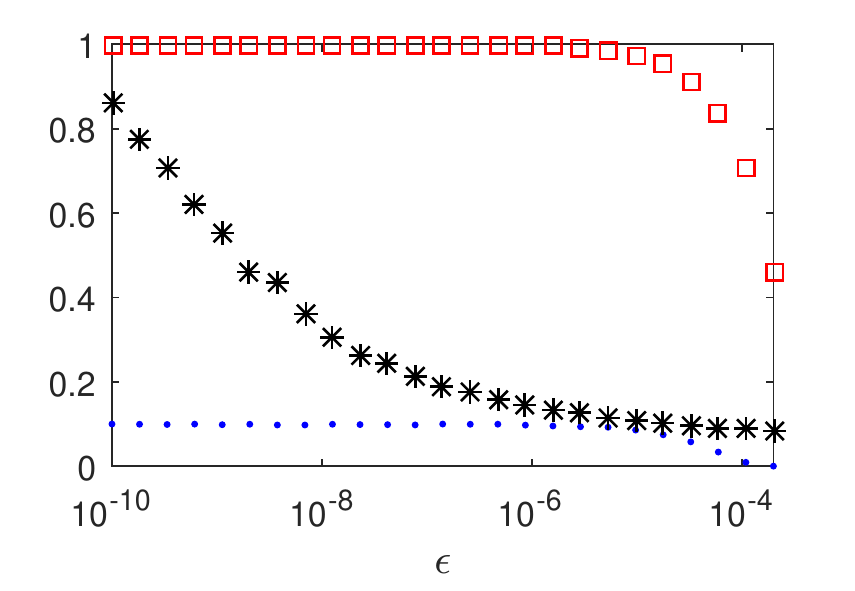}\hspace*{\fill}
\end{center}
\caption{\small Comparison of classical (clustered) and unclustered small-world networks in 2D random spatial networks.  All nodes have expected degree $k=20$.  Edges are placed using the distance kernel of~\eqref{eq:kernel} with $N_0= 10^5$ and $r_0=\sqrt{k/(\pi N_0)} = 8.0\times10^{-3}$.   
\textbf{(Left)} For $N=N_0= 10^5$ the expected number of nodes within a distance $r_0$ from a node is exactly $k$.  (Blue dots) Average local clustering coefficient, $c(\epsilon;N=N_0)$, normalized by $c(\epsilon=10^{-10};N=N_0)$. (Red squares) Average proximity to neighbor in the graph, i.e., one minus average geometric distance between neighbors, $d_{1}(\epsilon;N=N_0)$, normalized by $d_{1}(\epsilon=10^{-10};N=N_0)$. (Black asterisks) Average shortest path length in graph, $l(\epsilon;N=N_0)$, normalized by $l(\epsilon=10^{-10};N=N_0)$.
These networks show the classical small-world effect: as the fraction of long-range connections increases, local clustering persists, while the average shortest path length in the graph decreases rapidly. Note, however, that the local spatial proximity structure, i.e., the average proximity to neighbors, also persists for these classical small-world networks.
\textbf{(Right)} The same, but for $N =10^6= 10N_0$.  The normalization factors are the same as in the $N=N_0$ case.  These networks with $N \gg N_0$ show a conspicuous small-world effect: as the fraction of long-range connections increases, local spatial proximity persists, while the average shortest path length in the graph decreases rapidly. Note, however, that the local clustering is small for all $\epsilon$ for this new type of spatial small-world network, and is negligible in the large-$N$ limit.}
\label{fig:sw}
\end{figure*}
%------------------------

We first use the RSN model to generate  small-world networks satisfying the classical (clustered) definition.  We start with the unit torus $[0,1]\times[0,1]$ with periodic boundaries.  The networks we generate will have an average degree of $k$, such that nodes within some distance $r_0$ are very likely to be connected, and nodes of greater distance are very unlikely to be connected.  These are modeled after the Watts--Strogatz small-world networks~\cite{watts1998collective} with the number of local edges decreasing as the long-range connections increase so that the average degree remains fixed.

Specifically, we set $\kappa = k$ for all nodes and place $N=N_0:=k/(\pi r_0^2)$ nodes at random into the unit torus. Note that the unit torus has area 1. We assume $r_0<1/2$ so that disks of radius $r_0$ easily fit within the torus.  In particular this forces $\pi r_0^2<1$. The expected number of nodes within distance $r_0$ of a randomly chosen node is $k$, since $k/N_0=\pi r_0^2/1$.
We define
%------------------------
\begin{equation}
f(d_{uv}) = \begin{cases}
\frac{N_0}{k}\left[1-\epsilon\frac{1-\pi r_0^2}{\pi r_0^2}\right]& d_{uv} < r_0\\
\frac{N_0}{k}\epsilon & d_{uv} \geq r_0
\end{cases}.
\label{eq:kernel}
\end{equation}
%------------------------
%For $\epsilon =0$, this is the Random Geometric Graph.  
Equation (\ref{eq:p}) specializes to
%------------------------------------
\begin{equation}
p_{uv} =  \min\left(k \frac{f(d_{uv})}{N_0}, 1\right) = \frac{k f(d_{uv})}{N_0},
\label{eq:pspec}
\end{equation}
%------------------------------------
since $f(d_{uv})<N_0/k$ for small $\epsilon$ when $\pi r_0^2<1$.
%means that the term with $\epsilon$ is negative for $d_{uv}< r_0$ so $f(d_{uv})<N_0/k$.
Then 
%------------------------
\begin{equation}
p_{uv} = \begin{cases}
1- \epsilon\frac{1-\pi r_0^2}{\pi r_0^2} & d_{uv} <r_0\\
\epsilon & d_{uv}\geq r_0
\end{cases}.
\label{eq:puv}
\end{equation}
%------------------------
The expected number of nodes within distance $r_0$ is $\pi r_0^2 N_0=k$ and beyond distance $r_0$ is $(1-\pi r_0^2)N_0$.  The total number of expected neighbors is $k(1-\epsilon(1-\pi r_0^2)/(\pi r_0^2))+(1-\pi r_0^2)N_0 \epsilon=k$.

At $\epsilon=0$ all nodes within a distance $r_0$ are connected ($p_{uv}=1$) and no other nodes are joined.  This is a random geometric graph~\cite{penrose2003random}.  As $\epsilon$ increases, the number of long-range connections grows proportionally to $\epsilon$ with a corresponding decrease in the short-range connections.
%\JCM{I'm tempted to replace $\epsilon/\pi r_0^2$ with $\epsilon$}

A popular measure of clustering in a network is the local clustering coefficient $c_u$ of node $u$:
%---------------------------
\begin{align}
c_u= &\textrm{fraction of pairs of neighbors } \nonumber\\
  &\textrm{of node $u$ that are connected.} 
\end{align}
%---------------------------
This is also the number of triangles incident on $u$ divided by the total number of triangles that could be formed if all neighbors of $u$ were connected.
The average local clustering coefficient $c$ for the graph is
%---------------------------
\begin{equation}
c= \textrm{average}(c_u).
\end{equation}
%---------------------------
Both $c_u$ and $c$ lie between 0 and 1.  Values close to 0 indicate small clustering, and values
close to 1 indicate high clustering.

To measure the local \emph{spatial} structure in spatial networks we define two new quantities.
We first normalize all pairwise distances between nodes by the largest pairwise distance, such that all normalized
nodal distances lie between 0 and 1.
We then define the local proximity coefficient $d_{1,u}$ of node $u$ as
%---------------------------
\begin{align}
d_{1,u}= &1- \textrm{average geometric distance between} \nonumber\\ 
&\textrm{node $u$ and its neighbors in the network}.
\end{align}
%---------------------------
The local proximity coefficient $d_{1,u}$ of node $u$ lies between 0 and 1, with values close to 0 indicating low spatial proximity of the neighbours of $u$, and values close to 1 indicating high local spatial proximity.
Finally we define the average local proximity coefficient $d_{1}$ of the spatial network as
%---------------------------
\begin{equation}
d_1= \textrm{average}(d_{1,u}).
\end{equation}
%---------------------------
%Traditional views of small-world networks are defined by the property that the local character of the graph structure (high local clustering) is retained as long-range connections are added that quickly reduce the average shortest path length in the graph. We will demonstrate that for the processes we consider on spatial networks the occurrence of small-world effects as long-range connections are added and average shortest path lengths are rapidly reduced, is determined primarily by retention of the local \textit{spatial} character of the network (as indicated by high local spatial proximity), and that retaining clustering is not important for these small-world processes on spatial graphs.

For the network defined in \eqref{eq:puv} above, in the limit $\epsilon \to 0$, the clustering coefficient approaches $1-3\sqrt{3}/4\pi\approx0.5865$~\cite{dall2002random}, while the path length between two nodes $u$ and $v$ is at least their geometric distance divided by $r_0$: $d_{uv}/r_0$.  As $\epsilon$ increases, the clustering coefficient slowly decreases, while the path length between two nodes decreases much sooner (see the left hand side of Fig.~\ref{fig:sw}).  Thus this produces classical small-world networks.  As in the standard small-world models, we see that not only is the network clustered, but nearby nodes are preferentially connected, so there is spatial structure.

In Fig.~\ref{fig:sw} we plot three important network quantities for the network defined in \eqref{eq:puv}: the average shortest path length $l$ in the graph between any pair of nodes $u$ and $v$, the average local clustering coefficient $c$ of nodes $u$, and the average proximity $d_1$ to neighbors. These quantities are plotted as a function of $\epsilon$, which increases with the fraction of long-range connections in the graph. The classical small-world property is apparent on the left hand side: small increases in long-range connections rapidly decrease the average shortest path length in the graph, while the average local clustering coefficient is not affected by long-range connections until there are many more of them. As such, small-world networks will retain local connectivity while the network diameter is rapidly reduced, resulting in small-world effects.
We also plot the average proximity $d_1$ to neighbors, which plays a role similar to the clustering coefficient in the usual small-world models:  High local proximity is retained as long-range connections are added.

\subsection{Unclustered Small-World RSNs}

We now show that unclustered small-world networks share many of the same properties as classical small-world networks, but have no clustering.

To generate an unclustered small-world network, we mimic the process above, but rather than placing $N=N_0$ nodes into the unit torus, we place $N \gg N_0$ nodes.  We then place edges using the same distance kernel $f$.  This process increases the density of nodes by a factor of $N/N_0$, but the probability any two nodes are connected is reduced by the same factor, i.e., the probabilities in \eqref{eq:pspec} and \eqref{eq:puv} are multiplied by a factor $N_0/N$, because $\rho$ in \eqref{eq:p} now equals $N/V$. The expected number of nodes connected to $u$ is still $\kappa = k$, but there are now many nodes within distance $r_0$ from $u$ that are not connected to $u$.

The distribution of neighbor locations of a given node $u$ is the same under our classical small-world RSN networks as in this network, but the clustering coefficient is reduced by a factor of $N_0/N$.  Clustering is thus negligible in the large $N$ limit.  The resulting properties are shown on the right of Fig.~\ref{fig:sw} for $N=10 \, N_0$.

As in classical small-world networks,
the average shortest path length in the graph decreases rapidly as the fraction of long-range connections increases, but in these networks with $N =10^6 =10 N_0$ nodes the role of persistent large clustering is taken over by persistent close spatial proximity of nodes that are neighbors in the network.  We will show below that in these unclustered small worlds where the average shortest path length is small but spatial proximity of neighbors remains high, typical small-world effects arise.
%\JCM{A few conclusions from the figure.}

These unclustered small-world RSNs are much easier to work with.  They are locally tree-like, which will allow for analytic investigation of many spreading processes.  Below we will use this to study disease spread.

%------------------------

\subsection{Hop-and-Spread Dynamics in Small-World RSNs}

We have shown that unclustered and classical small-world networks share some structural properties.  We now show that the typical small-world aspects of the SIR epidemic process manifest themselves in a similar manner in both network types.

%------------------------
\begin{figure}[h]
% For faster compilation use PDF, in final version use PNG.
\includegraphics[width=\linewidth]{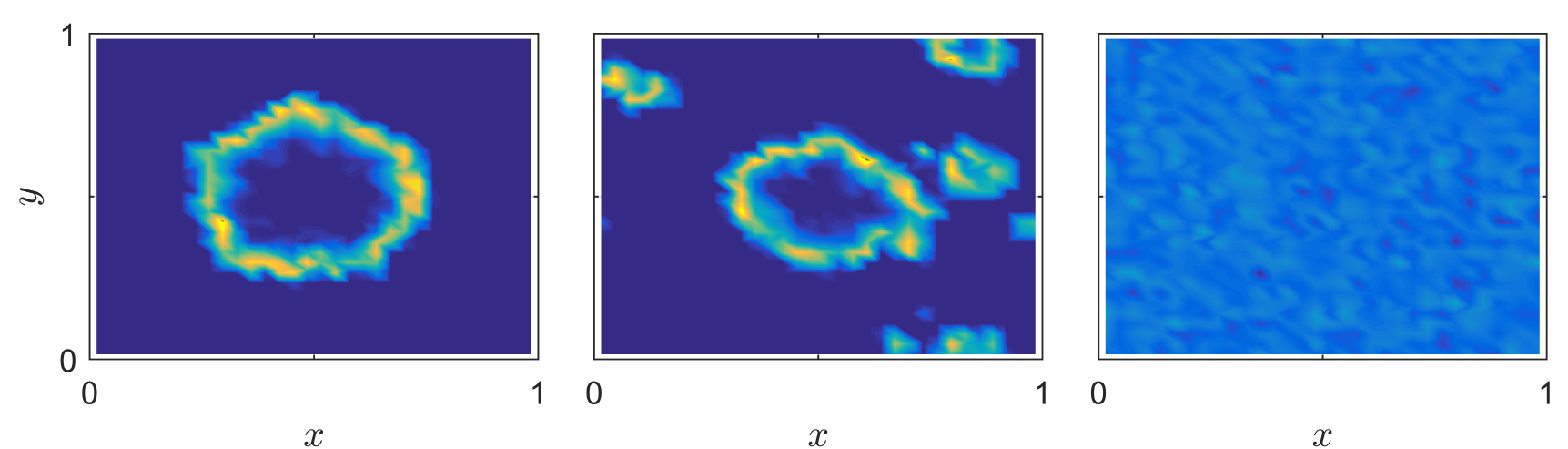}\\
\includegraphics[width=\linewidth]{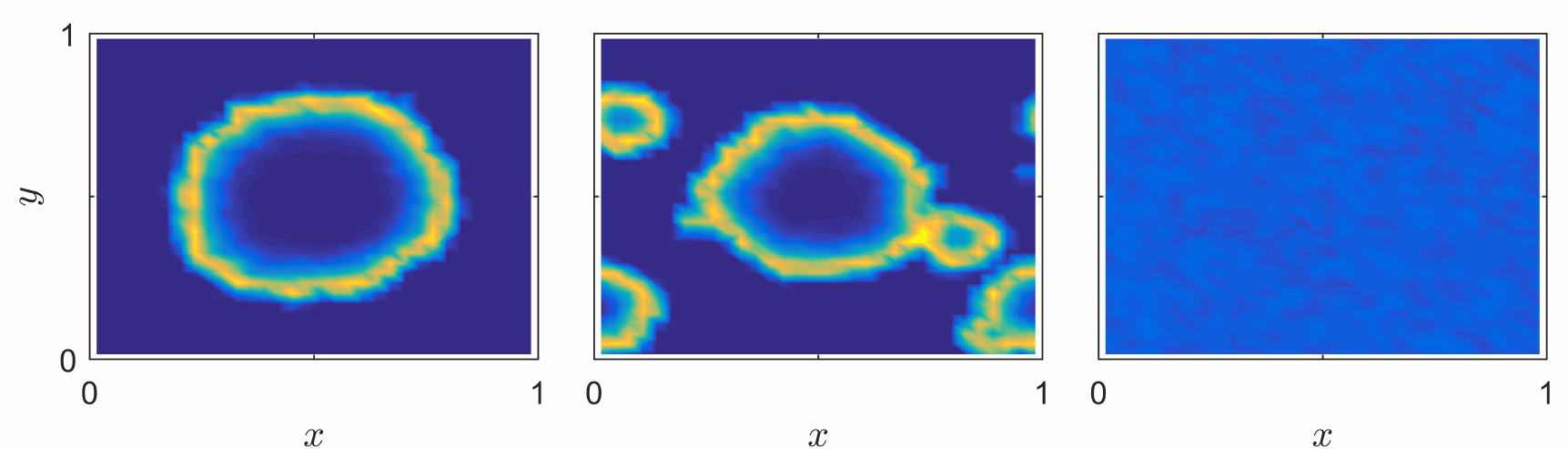}
\caption{\small Stochastic simulation of SIR disease dynamics on the 2D spatial networks of Figure \ref{fig:sw}, for $N=10^5$ (clustered, top) and $N=10^6$ (unclustered, bottom). Nodes transmit infection with rate $\gamma=3$ and recover with rate $\beta=1$. All nodes have expected degree $k=20$. The fraction of infected nodes, $I$, is shown, with a color scale that ranges from blue ($I=0$) to yellow ($I=0.5$).  The middle panels show that the small-world hop-and-spread dynamics occur in both the clustered and unclustered networks, i.e., small-world effects are demonstrated on spatial small-world networks without clustering.
\textbf{(Top)} $N=N_0=10^5$ nodes (high clustering for small $\epsilon$). The density of infected nodes is  shown for networks with (left) highly local spatial structure ($\epsilon=10^{-10}$), (middle) intermediate spatial structure ($\epsilon=10^{-7.25}$) with some long-range connections, and (right) no local spatial structure ($\epsilon=\pi r_0^2$, i.e., $f(d_{uv})=1$, uniform spatial kernel). Initial condition: the 100 nodes closest to (0.5,0.5) are initially infected. 
\textbf{(Bottom)}  $N = 10 N_0=10^6$ nodes (low clustering).  The density of infected nodes is shown for networks with (left) highly local spatial structure ($\epsilon=10^{-10}$), (middle) intermediate spatial structure ($\epsilon=10^{-8.25}$), and (right) no local spatial structure ($\epsilon=\pi r_0^2$, i.e., $f(d_{uv})=1$, uniform spatial kernel). Initial condition: the 1,000 nodes closest to (0.5,0.5) are initially infected. }
\label{fig:sw_disease}
\end{figure}
In an SIR epidemic, nodes begin susceptible and may be infected by infected neighbors (with rate $\beta$ per edge).  Eventually infected nodes recover (with rate $\gamma$ per node).  We will use stochastic simulation \cite{kiss:EoN} to study SIR disease on RSNs.

The simulated behavior in our clustered and unclustered small-world networks (Fig.~\ref{fig:sw_disease}) imitates our real-world observations:
\begin{itemize}
\item When there are few long-range connections ($\epsilon \to 0$), the disease spreads in a wave-like, coherent spatial pattern outwards from the source.
\item With an increased number of long-range connections, we see hop-and-spread dynamics.  The spread is locally wave-like until a long-range transmission occurs.  It then generates a new focal region.  
\item As the number of long-range transmissions increases, the disease spreads uniformly throughout the population.  Even if most connections are local and the network is highly clustered, sufficient long-range connections mean the epidemic behaves very similarly to how it would in a globally-mixed population.
\end{itemize}
The coherent spatial spread and globally-mixed regimes behave largely deterministically, but the intermediate hop-and-spread regime is dominated by stochastic effects. 

In a large enough domain (or equivalently, for small enough $r_0$), an epidemic spreading with any $\epsilon>0$ will eventually exhibit hop-and-spread dynamics.  How long the initial wave travels before the hop-and-spread dynamics begin depends on the time until the disease crosses a long-range connection.

% It is tempting to

This hop-and-spread behavior is naturally interpreted as being a consequence of the small-world structure of the network. However, as we see this behavior in both the clustered and unclustered networks (Fig~\ref{fig:sw_disease}), it is incorrect to think of it as requiring clustering.  Rather it requires the spatial proximity structure. The hopping behavior is clearly a consequence of the fact that nodes that appear far away are in fact close together in the network (see quote~(\ref{quote:close})), and the local wavelike propagation is facilitated by high spatial proximity of neighbors in the graph.  Although we argue the hop-and-spread dynamics is indeed a small-world effect, to do so we must widen the definition of small-world networks to include networks with this spatial structure even if they are unclustered.

We believe that many ``small-world'' effects on networks with built-in spatial structure are in fact a consequence of this spatial structure (high local spatial proximity of neighbors in the graph) rather than clustering.  RSNs allow us to investigate this in detail.
By tuning the distance kernel and density $\rho$, we can create small-world networks that are significantly
more flexible than the prototypical ring-based small-world networks with clustering.

We note however that for some processes, interactions with multiple neighbors may be required before a node changes status~\cite{centola:cascade,centola:weakness,centola:experiment}.  In such cases we believe clustering can play a role as it increases the likelihood that multiple neighbors will have a given status.  

%%%%%%%%%%%%%%%%%%%%%%%%%%%%%%%%%%%%%%%%%%%%%%
\section{Epidemic Spread on Random Spatial Networks: Analytical Models and Traveling Waves}
\label{sec:analytic}
%%%%%%%%%%%%%%%%%%%%%%%%%%%%%%%%%%%%%%%%%%%%%%

After considering the relation between RSNs and small-world networks in the previous section, we now demonstrate that RSNs are particularly well-suited for analytic study in the limit of large number of nodes.

We demonstrate this for SIR disease spread on RSNs.  We derive differential equations that govern the spatial dynamics of SIR disease on RSNs, and demonstrate the existence of nonlinear traveling wave solutions that retain their shapes.  We derive analytic wave speed expressions for the traveling waves, and numerically demonstrate a close match with the traveling wave structures arising in stochastic simulations of SIR disease on the RSNs.

%------------------------
\subsection{Analytic Governing Equations for SIR Disease on RSNs}
%------------------------
We consider differential equations that accurately describe SIR disease dynamics on RSNs.  We have noted that in the high-density limit RSNs have the crucial property of being locally tree-like, which allows us to apply analytic tools that have been applied to non-spatial networks that are locally tree-like. 

As we demonstrate for SIR dynamics, we can derive analytic equations governing the deterministic dynamics (that is, assuming that the number of nodes is sufficiently large for stochastic effects to become negligibly small).  We base our method on analytic techniques derived for Chung-Lu or Configuration Model networks~\cite{miller:ebcm_overview}.  Similar adaptations will apply for analytic models of other processes.

We work in the $\rho \to \infty$ limit.  We define $S(\vec{x},t)$, $I(\vec{x},t)$ and $R(\vec{x},t)$ to be the probability a node at location $\vec{x}$ and time $t$ would be susceptible, infected or recovered.  We assume that the disease is introduced to the nodes at time $t=0$ with a probability $I(\vec{x},0)$ that depends only on location.  We assume all other nodes are susceptible $S(\vec{x},0) = 1- I(\vec{x},0)$ and introduce the variable $\Theta(\vec{x},t)$ which is the probability an edge belonging to node $u$ at position $\vec{x}$ has not transmitted infection to $u$ by time $t$. Applying techniques from~\cite{miller2012edge} extended to spatial networks (see SI), in particular assuming that stochastic fluctuations are negligible, yields
%------------------------------------
\begin{subequations}
\label{sys:PIDE}
\begin{align}
\pd{}{t}{\Theta}(\vec{x},t) &= - \beta \Theta(\vec{x},t) +
\gamma(1-\Theta(\vec{x},t)) \nonumber \label{eq:dtheta}\\ 
&\quad+ \beta
\frac{\int_{V}S(\hat{\vec{x}},0)\Psi'(\Theta(\hat{\vec{x}},t)) f(|\hat{\vec{x}}-\vec{x}|) \,
  \mathrm{d}\hat{\vec{x}}}{\ave{\kappa}}\\
S(\vec{x},t) &= S(\vec{x},0)\Psi(\Theta(\vec{x},t)),\\ 
%\dot{R}
\pd{}{t}{R}(\vec{x},t) &= \gamma (1-S(\vec{x},t)-R(\vec{x},t))\\
I(\vec{x},t) &=1-S(\vec{x},t)-R(\vec{x},t) \label{eq:I}
\end{align}
\end{subequations}%------------------------------------
with $\Psi(\Theta) =
\int_0^\infty e^{-\kappa(1-\Theta)} P(\kappa) \,
\mathrm{d}\kappa$. Our initial conditions are $\Theta(\vec{x},0)=1$ and $R(\vec{x},0)=0$, with $S(\vec{x},0) = 1- I(\vec{x},0)$.
%and we impose an initial infection with the initial infected fraction $I(\vec{x},0)$, and initial susceptible fraction $S(\vec{x},0)=1-I(\vec{x},0)$.  

The equation for $\Theta$ is a non-local evolution equation.  The non-local effects are captured through the convolution integral.  To emphasize that the non-local interactions are captured by an integral, we refer to this as a partial integro-differential equation (PIDE): the integral is over space and the derivative is with respect to time.
What is new and significant compared to the results from~\cite{miller2012edge} for non-spatial networks, is that we obtain an evolution equation for ${\Theta}(\vec{x},t)$ that captures intricate spatial effects on the RSN in the integral term.

We expect the equation for $\Theta$ to be similar to the Fisher--KPP equation for a spreading population~\cite{fisher1937wave,KPP} $u_t = ru(1-u/K) + D u_{xx}$, with the spatial integral playing the role of the nonlinear and diffusion terms.  The spatial integral captures the network's structure by including non-local interactions through $f$ and the degree distribution through $\Psi$.  In the SI we derive the Fisher--KPP equation as an approximation of~\eqref{eq:dtheta}.  Other non-local versions of the Fisher--KPP equation have been studied~\cite{coville2005propagation,berestycki2009non}, including some for disease spread~\cite{diekmann1978thresholds,diekmann1979run}.  These are based on a mass-action mixing assumption and involve a convolution of a distance kernel with $u$ directly rather than a nonlinear function of $u$. 

Since stochastic simulations on networks are hard to analyze and understand, this explicit analytic equation provides a powerful tool to study SIR dynamics on random networks with versatile spatial structure, fully taking into account the distribution of expected degrees and the distance kernel. We now demonstrate how this equation allows us to identify and characterize traveling waves solutions on RSNs that match stochastic simulation results.

%In this paper, we demonstrate this for SIR disease spread in the large density limit.  We can adapt the models of~\cite{miller:ebcm_overview} to get
%------------------------
%\begin{align*}
%\pd{}{t} \Theta(\vec{x},t) &= - \beta \Theta(\vec{x},t) + \gamma(1-\Theta(\vec{x},t))\\
%&\quad + \beta \frac{\int_{V}\Psi'(\Theta(\hat{\vec{x}},t)) f(\hat{\vec{x}}-\vec{x}) \, \mathrm{d}\hat{\vec{x}}}{\Psi'(1)}\\
%S &= \Psi(\vec{x},\Theta(\vec{x},t)) \\
%I &= 1-S-R\\
%\dot{R} &= \gamma I
%\end{align*}
%------------------------
%Here
%------------------------
%\[
%\Psi(\vec{x},y) = \int_0^\infty S(\vec{x},0) e^{-\kappa(1-y)} P(\kappa) \, \mathrm{d}\kappa
%\]
%------------------------
%and $S(\vec{x},0)$ is the probability a random node at location $\vec{x}$ is susceptible at time $0$.

%\HDS{need to explain that this is partial integro-differential equation, explain what all the symbols mean, give some high-level insight into how this can be derived, referring to the SI; point to importance as extension of Fischer; explain power and significance of obtaining explicit equation to describe SIR on RSN, taking into account degree distribution and distance kernel}

%------------------------
\subsection{Traveling Waves and Correspondence between Stochastic SIR Simulation and PIDE Model}
%------------------------
%\HDS{Discussion of Fig. 6. Hans can write this part (text below from proposal)}

We have identified nonlinear traveling waves in stochastic simulations of SIR dynamics on RSNs. These waves retain their shapes as they evolve in time. We first compare stochastic SIR simulations of traveling waves on RSNs with numerical simulations of System~(\ref{sys:PIDE}) to demonstrate that these equations closely describe the stochastic dynamics in the limit of large $N$. In the next subsection we then derive existence conditions and wave speed properties of the traveling waves for System~(\ref{sys:PIDE}), illustrating the analytical power of our approach.

We consider stochastic simulations on one-dimensional (1D) and two-dimensional (2D) RSNs generated using an algorithm based on the linear-time algorithm of~\cite{miller2011efficient} for Chung--Lu networks (see SI). We simulate epidemic spread starting from a localized initial condition using a stochastic simulation algorithm from~\cite{kiss:EoN}.

\begin{figure}[h]
\includegraphics[width=\linewidth]{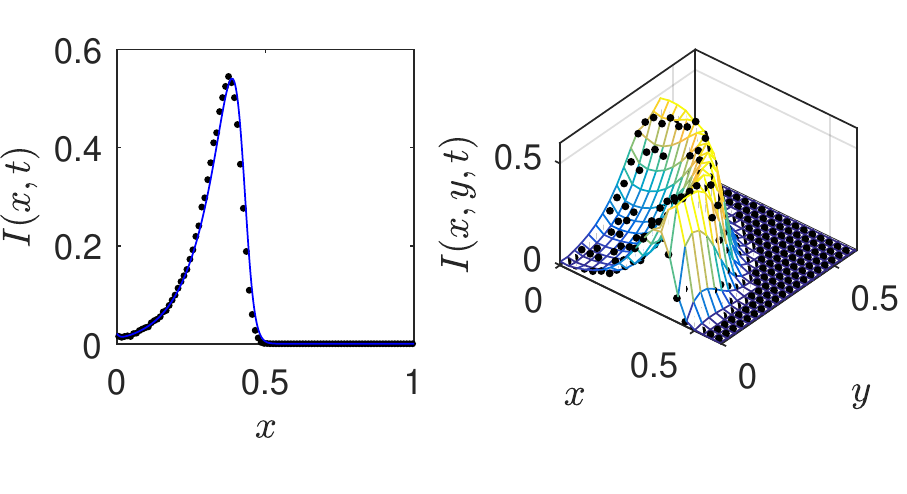}
\caption{\small Comparison of stochastic simulation of SIR disease dynamics with numerical solution of analytical PIDE, on 1D and 2D RSNs. In both panels, results of stochastic simulation on networks with $N=10^6$ nodes are given by black dots. (Left) 1D PIDE solution is given by solid blue line (spatial resolution $\Delta x = 10^{-4}$). (Right) 2D PIDE solution is given by mesh surface (spatial resolution $\Delta x = 1/512$). Networks in both panels are generated using kernel $f(r) = \phi_{0,0.01}(r)$, where $\phi_{\mu,\sigma}(x)$ is the probability density function for the normal random variable with mean $\mu$ and standard deviation $\sigma$. All nodes have expected degree $\kappa=20$. Initial conditions: (left) 10\% of nodes in the interval $[0,0.01)$ are initially infected, and (right) 10\% of nodes in the square $[0,1/32)\times[0,1/32)$ are initially infected.
% t=1.76
}
\label{fig:traveling}
\end{figure}

Figure~\ref{fig:traveling} shows travelling waves revealed by the stochastic simulations in 1D and 2D.  In 1D, these retain their shape as they propagate. Note that the waves observed previously in Fig.~\ref{fig:sw_disease}
are also traveling waves, as can be seen in SI movies 1--6. %\JCM{DOUBLE CHECK MOVIES}
Figure~\ref{fig:traveling} also shows numerical PIDE solutions of System~(\ref{sys:PIDE}) for the 1D and 2D network problems, demonstrating good agreement between PIDE solution and stochastic simulations on the 1D and 2D RSNs with Gaussian spatial kernel.
%The shapes and amplitudes of the waves agree well, but we find there is some difference in the stochastic simulation wave speed and the wave speed from the numerical solution of the PIDE. For low PIDE grid spacing and/or low node density in the stochastic simulations (as in Figure~\ref{fig:traveling}), Fig.\ \ref{fig:waveSpeed} shows that wave speeds may indeed differ. (In Fig.~\ref{fig:traveling} we have shifted the waves to align them, to correct for these low-resolution effects.)

%Close inspection reveals that the wave in the numerical PIDE solution travels somewhat faster.  This will be discussed in more detail in the following section.

%------------------------
%------------------------

%------------------------
\subsection{Analytic Properties of Traveling Waves on Random Spatial Networks}
%------------------------
The PIDE formulation provides a powerful technical tool to derive precise quantitative insight in the spread of SIR disease on RSNs. As a major application, we can derive properties of the traveling waves that were identified in Figs.\ \ref{fig:sw_disease}
%, \ref{fig:unclustered_sw_disease} 
and \ref{fig:traveling}. 
In particular, we derive an explicit expression for the wave speed of the 1D traveling wave, and identify conditions on the spatial decay of the kernel for the traveling wave to exist. This wave has a ``pulled front'': That is, its speed is set by the nodes in the leading edge.
%------------------------
%\subsubsection{Wave speed for traveling wave}
%------------------------
%\

%\JCM{Should probably switch $f$ to being an even function and drop the absolute values.  We haven't been consistent.}
% I think I have added the absolute values everywhere where necessary ... (Hans)

We derive the wave speed assuming that the domain $V$ is the real line.  We can write $\Theta(x,t) = \theta(\xi(x,t))$ where $\xi(x,t) = x-ct$ and $c$ is the wave speed. We will assume it is traveling from left to right.  The wave travels into a region where $S(x,t)\approx 1$, \ $I(x,t) \ll 1$ and $R(x,t) \ll 1$, and very little transmission has occurred, so ahead of the traveling wave $\Theta \approx 1$. We assume $\xi$ is in the leading edge and write $\Theta(x,t) = \theta(\xi(x,t))=1-\epsilon h(\xi(x,t))$.

In this leading edge of the wave $\epsilon h(\xi) \ll 1$, while in the bulk of the wave, $\epsilon h(\xi)$ may be comparable to $1$.  We focus on the behavior in the leading edge of the wave and transform the equation for $\pd{}{t}\Theta$ (\eqref{eq:dtheta}) into an equation for $h$ by expanding $\Psi'(\theta(\xi))$ as a Taylor Series about $\theta=1$:
\[
\Psi'(\theta) = \Psi'(1) - \epsilon h(\xi) \Psi''(1) + \frac{\epsilon^2 h(\xi)^2}{2}\Psi'''(1) + \cdots
\]
Note that $\Psi^{(n)}(1) = \ave{\kappa^n}$, that is, the $n$th derivative of $\Psi$ evaluated at $1$ is the average of the $n$th power of $\kappa$.

Substituting this into the integral (and taking $S(x,0)=1$ ahead of the wave), we arrive at
\begin{align*}
\epsilon c h'(\xi) &= -\beta(1-\epsilon h(\xi)) + \epsilon \gamma h(\xi) \\
&+\beta \frac{\int_{-\infty}^\infty [\Psi'(1) - \epsilon h(\hat{\xi}) \Psi''(1) + \order(\epsilon ^2h(\hat{\xi})^2)]f(|\hat{\xi}-\xi|) \mathrm{d}\hat{\xi}}{\ave{\kappa}}
\end{align*}
Because $\Psi'(1)=\ave{\kappa}$, the $\order(1)$ terms cancel.  We neglect the $\order(\epsilon^2 h(\hat{\xi})^2)$ terms by arguing that if $|\hat{\xi}-\xi|$ is not large then $\epsilon^2h(\hat{\xi})^2$ is small (at the leading edge), and if $|\hat{\xi}-\xi|$ is large then $f(|\hat{\xi}-\xi|)$ is small (away from the leading edge).

This leaves the linear homogeneous equation for $h$
\[
ch'(\xi) = (\beta+\gamma)h(\xi) + \beta \frac{\ave{\kappa^2}}{\ave{\kappa}}\int_{-\infty}^\infty h(\hat{\xi})f(|\hat{\xi}-\xi|) \, \mathrm{d}\hat{\xi}
\]
We anticipate  $h(\xi) \approx e^{-\alpha \, \xi}$  for some $\alpha$, yielding:
\[
-c\alpha = (\beta+\gamma) + \beta \frac{\ave{\kappa^2}}{\ave{\kappa}}\int_{-\infty}^\infty e^{-\alpha (\hat{\xi}-\xi)} f(|\hat{\xi}-\xi|) \, \mathrm{d}\hat{\xi}
\]
Setting $u = \hat{\xi}-\xi$ the integral becomes $\int_{-\infty}^\infty e^{-\alpha u} f(|u|)$.  This is the bilateral Laplace transform of $f(|x|)$, which we denote $\Lap{}[f](\alpha)$.  Performing some further algebra yields
\begin{equation}
\frac{c}{\beta+\gamma} = -\frac{1}{\alpha} + \Ro \frac{\Lap{}[f](\alpha)}{\alpha} \label{eq:alpha_c}
\end{equation}
where $\Ro = \beta\ave{\kappa^2}/(\beta+\gamma)\ave{\kappa}$ (this is the basic reproductive number of the SIR disease on the network, which is the typical number of infections caused by an infected individual early in an epidemic~\cite{andersonMay}).
%\HDS{Joel, perhaps remind readers of the definition of basic reproductive number, with a reference.}

There are infinitely many solutions $\alpha,c$. Following results for the Fisher--KPP equation we expect that the observed solution has the minimum wave speed.  Setting $\pd{c}{\alpha}=0$ and performing some algebra shows that this occurs when
\begin{equation}
\alpha\Lap{}[xf(|x|)](\alpha) + \Lap{}[f](\alpha) = \frac{1}{\Ro}
\label{eqn:find_alpha}
\end{equation}
We can solve this equation to find $\alpha$.

Finally, substituting~\eqref{eq:alpha_c} into~\eqref{eqn:find_alpha} gives
\begin{equation}
c = -\beta\frac{\ave{\kappa^2}}{\ave{\kappa}} \Lap{}[xf(|x|)](\alpha)
\label{eqn:find_c}
\end{equation}

If $f$ does not decay at least exponentially fast, then these Laplace transforms do not exist.  Thus we infer that if $f$ does not decay at least exponentially fast, then there is no coherent traveling wave solution in the $\rho\to\infty$ limit.  In practice, for a finite population if the long tail is observed by the transmissions, we see hop-and-spread dynamics, while if the long tail is not sampled by the transmission we may still see traveling wave behaviors.

%\HDS{Joel, I've read and edited until this point. This has to be amended by some of the stuff below (but most should be commented out, I guess). Also, we should make sure to comment further on the `mismatch' that was mentioned in the previous subsection.}

%------------------------
\begin{figure}[h]
\includegraphics[width=\linewidth]{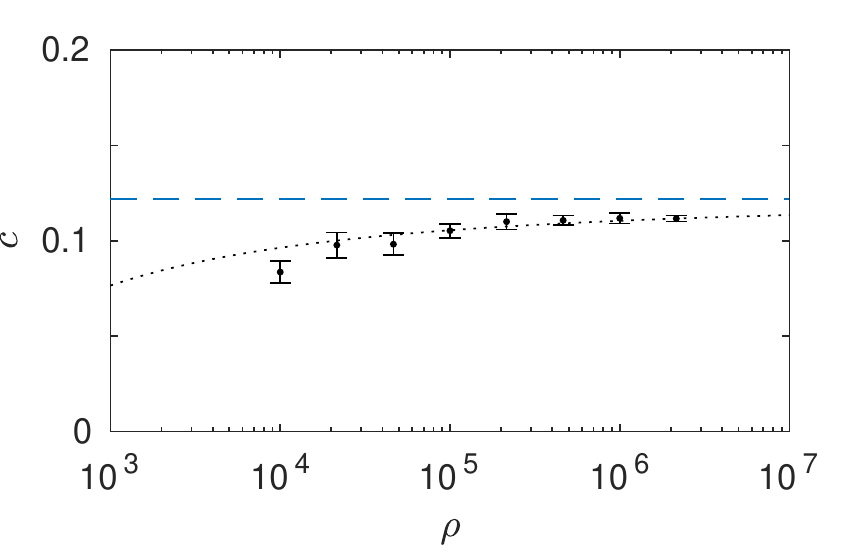}
\caption{\small Comparison of wave speed for SIR disease dynamics observed in stochastic simulations (small black circles) and analytic prediction (dashed cyan line). For the stochastic simulations, we generate $n_{rep} = 25$ 1-D spatial networks for each node density $\rho$ using a distance kernel $f(|x|) = \phi_{0,0.01}(|x|)$, where $\phi_{\mu,\sigma}(x)$ is the probability density function for the normal random variable with mean $\mu$ and standard deviation $\sigma$. All nodes have expected degree $\kappa=10$. For each network we realize one SIR simulation with disease parameters $\beta=1$ and $\gamma=3$. The black circles show the average wave speed resulting from 25 network realizations (vertical bars indicate 95\% confidence intervals).  
The dotted black curve represents the expected convergence behavior of $c^*- K/ln^2\rho$ to the analytically predicted wave speed $c^*$, where the constant $K$ is obtained by fitting the curve to the rightmost black circle, which has the smallest error bar since it corresponds to the largest node density $\rho$.}
\label{fig:waveSpeed}
\end{figure}
%------------------------

%------------------------
\begin{figure}[h]
\includegraphics[width=\columnwidth]{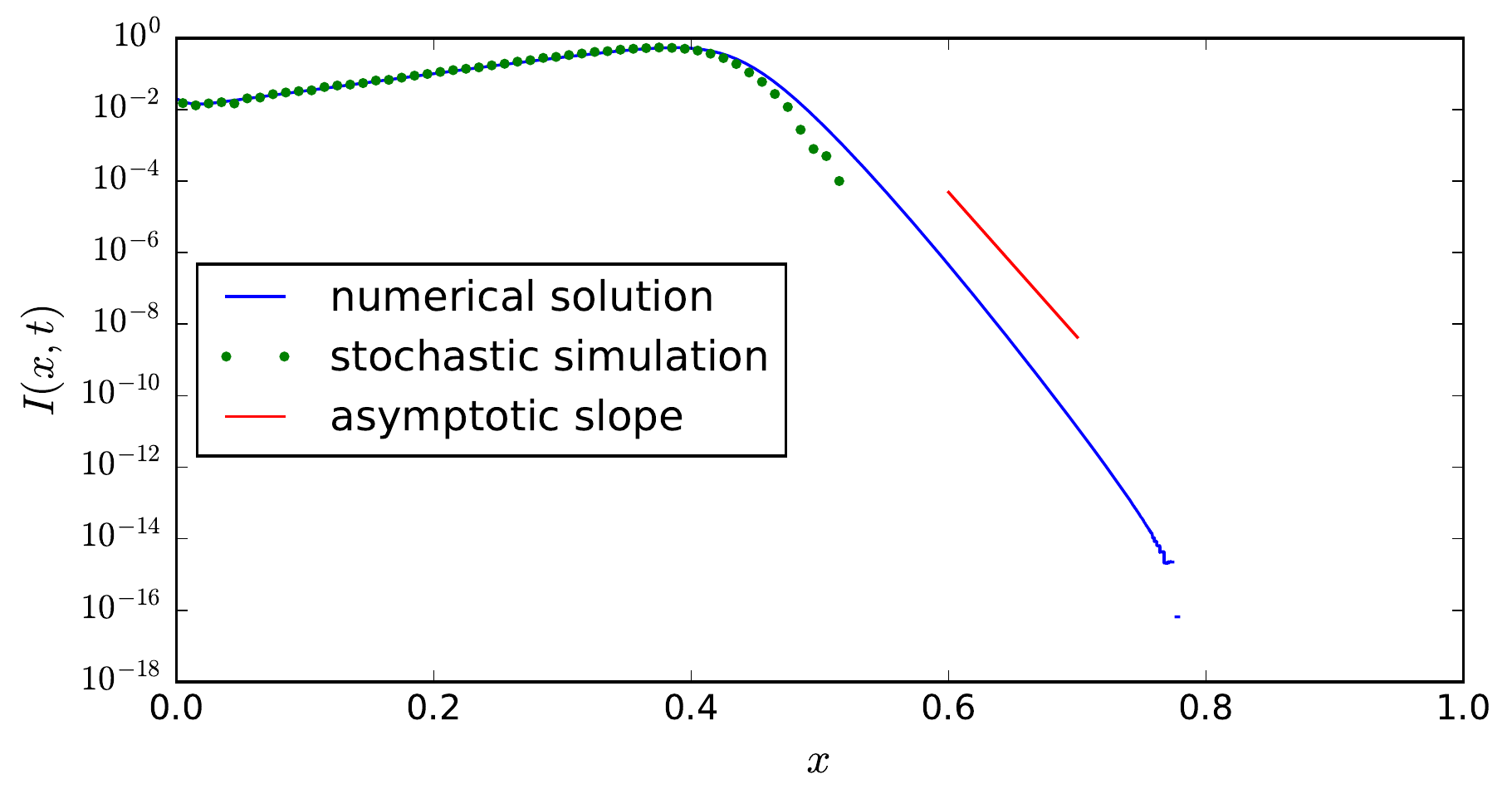}
\caption{\small A comparison of stochastic simulation and numerical PIDE solution on a log-scale, using the 1D solutions from Fig.~\ref{fig:traveling}.  Both stochastic and numerical solutions experience leading edge truncation.  For the stochastic simulation it is due to the finite number of nodes, while for the numerical solution it is due to numerical precision ($\sim 10^{-16}$).  The slope of the leading edge is close to the asymptotic prediction of $-\alpha$.}
\label{fig:num-exp}
\end{figure}
%------------------------

Figure \ref{fig:waveSpeed} compares the analytically predicted wave speed $c$ with wave speeds obtained from stochastic simulations.  

The small black circles of Fig.~\ref{fig:waveSpeed} give the wave speed observed in stochastic simulations. They converge to the analytic prediction (cyan dashed), but the convergence is very slow. Due to finite density, the exponentially decaying front is eventually truncated at its leading edge as shown in Fig.~\ref{fig:num-exp}. This truncation slows the wave because the foremost infected nodes play a large role in setting the wave speed.  This has been analyzed for similar fronts in other stochastic systems, for which the leading order error in the wave speed decays proportionally to $1/\ln^2 \rho$~\cite{panja2004effects}. More nodes are needed to improve the fit, as seen in Fig.~\ref{fig:waveSpeed}.
Fig.~\ref{fig:num-exp} confirms that the stochastic simulation front and the numerically calculated PIDE front are nearly exponential with slope close to the predicted $-\alpha$.
For the stochastic simulation (green dots), local densities in the exponentially decaying profile smaller than about 10$^{-4}$ cannot be represented because there are insufficient points in the simulation (the local densities effectively drop down to zero to the right of $x\sim 0.5$, and these zero values end up outside the range of the vertical logarithmic axis of the figure). By having larger $\rho$, more of the leading edge is observed resulting in wave speeds closer to the analytic prediction (Fig.~\ref{fig:waveSpeed}). Fig.~\ref{fig:num-exp} also shows that, in the numerical PIDE simulations, the exponential profile is truncated when the density of infected individuals approaches machine accuracy.

%------------------------
\subsection{Epidemic Final Size}
%------------------------
We can predict the final size of an epidemic in a large population.  As $t \to \infty$ the system approaches an equilibrium.  By setting $\pd{}{t}\Theta = 0$, we have
\[
\Theta(\vec{x},\infty) = \frac{\gamma}{\beta+\gamma} + \frac{\beta}{\beta+\gamma} \frac{S(\hat{\vec{x}},0)\int_V  \Psi'(\Theta(\hat{\vec{x}},\infty)) f(|\hat{\vec{x}}-\vec{x}|) \, \mathrm{d}\hat{\vec{x}}}{\ave{\kappa}}
\]
If $\vec{x}$ is far from the point of introduction, or the introduction is so small that we can approximate $S(\vec{x},0)$ as $1$ everywhere, we can treat $\Theta(\vec{x},\infty)$ as spatially homogeneous.  Then 
\begin{align}
\Theta &= \frac{\gamma}{\beta+\gamma} + \frac{\beta}{\beta+\gamma} \frac{\Psi'(\Theta)}{\ave{\kappa}} \, ,
\label{eqn:theta_fs}\\
S &= \Psi(\Theta) \, .
\end{align}
$\Theta=1$, \ $S=1$ is always a solution, but if there is a solution $\Theta$ between $0$ and $1$, it is the appropriate one for an epidemic.  It exists precisely when $\Ro>1$.  This is the same relation as for a random Chung-Lu network without spatial structure~\cite{miller:ebcm_overview}.    Interestingly, this means the epidemic final size in RSNs is independent of the distance kernel and depends only on disease properties and the degree distribution.
%typical degrees.

%%%%%%%%%%%%%%%%%%%%%%%%%%%%%%%%%%%%%%%%%%%%%%
\section{Discussion and Conclusion}
 \label{sec:conclusion}
%%%%%%%%%%%%%%%%%%%%%%%%%%%%%%%%%%%%%%%%%%%%%%

%------------------------
%\subsection{Possible Generalizations for the RSN Model}
%------------------------

The RSN model defined by~\eqref{eq:p} is versatile as it allows flexible degree distributions and distance kernels.
There are some further generalizations that were not considered in this paper but offer compelling prospects for
building realistic networks models that remain analyzable.

\begin{figure}
%\missingfigure{need to reproduce Rambaut et al's fig}
%\includegraphics[width=\linewidth]{FigS3a_kernel}
\centering
\includegraphics[width=.8\linewidth]{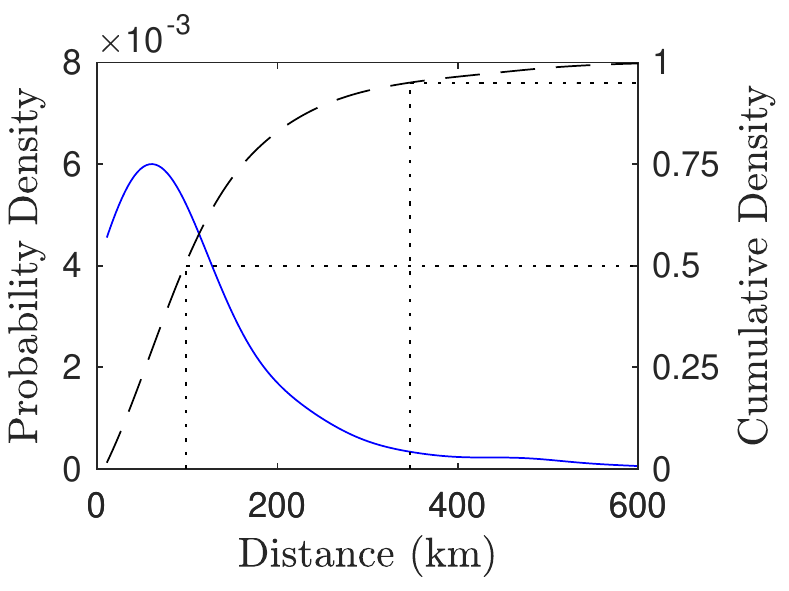}
\caption{\small Observed hop distances inferred from viral genome sequencing for the 2013-2016 Ebola outbreak in West-Africa. Data courtesy of A. Rambaut \cite{dudas2016virus}. The estimated probability density is obtained from raw binned data by Kernel Density Estimation using a Gaussian kernel function.  The dotted lines denote the 50th and 95th percentiles.}
\label{fig:Ebola_kernel}
\end{figure}

An obvious generalization is to allow $\rho(\vec{x})$ to describe an inhomogeneous spatial density of nodes.
This could, for example, be used to model different population densities in cities and rural areas in the context of realistic spatial disease spreading models, or different densities of neurons in different parts of the brain. Similarly, instead of letting the distance kernel depend on nodal distances $d_{uv}$, one can consider more general kernels $f(\vec{x}_u,\vec{x}_v)$ that directly depend on the coordinates of the nodes. For example, this could model people living in cities preferentially connecting to people in the same and other cities, while connections with and between rural individuals could follow different patterns. This type of modeling is especially compelling in the era of big data where real data may be used to build analyzable spatial random network models that faithfully mirror real-world spatial networks.
For example, \cite{balcan2009multiscale} studies the interplay between short-scale commuting flows and long-range airline traffic in shaping the spatiotemporal pattern of a global epidemic, and \cite{dudas2016virus} found that during the 2013-2016 Ebola outbreak viral lineages moved according to a classic ``gravity'' model, with more intense migration between larger and more proximate population centers. 
In this vein, Fig.~\ref{fig:Ebola_kernel} shows the observed hops seen from virus genome sequencing for the 2013-2016 Ebola outbreak, which can be built into RSN models.
Using empirical data or inferred spatial connectivities, all these kinds of effects can be incorporated in our RSN models of~\eqref{eq:p}, with clear potential for highly realistic models that retain the analytical tractability of the approach. 

%--------------------------------------------------
%--------------------------------------------------

The RSN class is also amenable to graph theoretic analysis. For example, it would be of great interest to investigate thresholds for the existence of a giant component for RSNs with various distance kernels. This was done with great success for non-spatial random graphs with given degree sequences by Reed and co-workers \cite{molloy1995critical,joos2016determine}. RSNs add a missing spatial component to this setting and are promising because they do so in a way that remains analyzable.

In summary, we propose a class of random spatial networks (RSNs) that intrinsically incorporates spatial structure in a way that crucially remains analytically tractable. We use RSNs to describe a new generalization of small-world networks without clustering, where the role of large clustering is taken over by close spatial proximity of nodes that are neighbors in the graph. We believe that many small-world effects on networks with built-in spatial structure are in fact a consequence of this spatial structure rather than clustering. We observe nonlinear traveling waves on RSNs in the context of SIR disease propagation, and, through analytical derivation of new governing partial integro-differential equations on graphs, we are able to precisely characterize these waves analytically. This is the first quantitatively accurate analytical description of nonlinear traveling waves on graphs with spatial structure. 

Our SIR disease model on spatial graphs extends analytical approaches that were previously only available to describe dynamics at the level of populations or on networks without spatial structure, to the real-life case of networks characterized by important spatial structure. This provides a theoretical modeling framework, for example, for recent observations of how epidemics like Ebola evolve in modern connected societies, with long-range connections seeding new focal points from which the epidemic locally spreads in a wavelike manner. There are many possibilities for further applications in disease modeling. For example, the SIR traveling wave models can be applied to realistic simulations of historic epidemics such as the plague in medieval Europe, incorporating in a precise manner the geography and estimated historical population density maps. Further potential applications include models for a diversity of areas such as neuronal networks in the human brain, spread of animal and plant species in fragmented habitats, and propagation of civil unrest, or public health epidemics such as obesity and smoking, in human societies.

%%%%%%%%%%%%%%%%%%%%%%%%%%%%%%%
%%%%%%%%%%%%%%%%%%%%%%%%%%%%%%%
%%%%%%%%%%%%%%%%%%%%%%%%%%%%%%%
%\clearpage

\subsection*{Acknowledgment}
JCM was funded by the Global Good Fund through the Institute for Disease Modeling and by a Larkins Fellowship from Monash University.  JLK was funded by an AMSI Vacation Research Scholarship. JL acknowledges funding from the Natural Sciences and Engineering Research Council of Canada.
%\JCM{JL and HDS funding}

% \pnasbreak splits and balances the columns before the references.
% If you see unexpected formatting errors, try commenting out this line
% as it can run into problems with floats and footnotes on the final page.
%\pnasbreak

% Bibliography
%%%%%%%%%%%%%%%%%%%%%%%%%%%%%%%%%%%%%%%%%%%%%%
{
\scriptsize \bibliographystyle{plain}
\bibliography{spatial_networks}
}
%%%%%%%%%%%%%%%%%%%%%%%%%%%%%%%%%%%%%%%%%%%%%%
%%%%%%%%%%%%%%%%%%%%%%%%%%%%%%%%%%%%%%%%%%%%%%

%%%%%%%%%%%%%%%%%%%%%%%%%%%%%%%%%%%%%%%%%%%%%%%%
%%%%%%%%%%%%%%%%%%%%%%%%%%%%%%%%%%%%%%%%%%%%%%%%
%%%%%%%%%%%%%%%%%%%%%%%%%%%%%%%%%%%%%%%%%%%%%%%%
%%%%%%%%%%%%%%%%%%%%%%%%%%%%%%%%%%%%%%%%%%%%%%%%

\clearpage

%%%%%%%%%%%%%%%%%%%%%%%%%%%%%%%%%%%%%%%%%%%%%%%%
%%%%%%%%%%%%%%%%%%%%%%%%%%%%%%%%%%%%%%%%%%%%%%%%
%%%%%%%%%%%%%%%%%%%%%%%%%%%%%%%%%%%%%%%%%%%%%%%%
%%%%%%%%%%%%%%%%%%%%%%%%%%%%%%%%%%%%%%%%%%%%%%%%

\section*{Supporting Information}

In this Supporting Information, we first explain in detail how several important existing random network classes can be obtained as special cases of our Random Spatial Network (RSN) class. We then provide some details on an algorithm that can generate RSNs efficiently. We summarize how analytic equations can be formulated for SIR disease on networks, using the general approach from \cite{volz:cts_time, miller:ebcm_overview, miller:initial_conditions}, but extended to the spatial setting. This explains how we obtain the partial integro-differential equations that govern SIR disease spread on RSNs as given in the main text.  Finally we show that the analytic equations can be reduced to the Fisher--KPP equation if we make strong assumptions.

\subsection{Special cases of RSNs}
\label{sec:special-cases}
Many existing models (with and without spatial structure) arise as special cases of RSNs by a particular combination of choices of the distribution of expected degrees and the distance kernel, or by placing the nodes at lattice points.
%\JCM{I lean towards moving this section to supplement and simply listing these networks here.}
\subsubsection{Geometric Inhomogeneous Random Graphs}

The Geometric Inhomogeneous Random Graph model of~\cite{bringmann2016geometric} places nodes uniformly in some space, gives each node $u$ a weight $w_u$, and then assigns edges between two nodes $u$ and $v$ with probability proportional to $w_uw_v$ and inversely proportional to some power of their distance $d_{uv}$.
We recover this model by taking the distance kernel $f$ to be some negative power of the distance between two nodes.  Thus the Geometric Inhomogeneous Random Graphs have a distance kernel that decays algebraically.  We show in the main text that significantly different dynamic behaviors can emerge when the distance kernel decays exponentially or faster. 

%------------------------
\subsubsection{Waxman Graphs}
%------------------------
The Waxman graph model~\cite{waxman1988routing} places nodes uniformly in a 2-dimensional rectangle.  An edge is placed between nodes $u$ and $v$ with probability $\beta \exp(-d_{uv}/L\alpha)$ for constants $\alpha$ and $\beta$ and maximum distance $L$.  
We can recover this model by setting $\kappa_u = \ave{\kappa}$ for all $u$ and choosing a decaying exponential as the distance kernel.

%------------------------
\subsubsection{Random Geometric Graph}
%------------------------
In a Random Geometric Graph, nodes are placed uniformly at random into  $V$, and any nodes whose distance is less than some value $r_0$ are joined together~\cite{penrose2003random}.  

We can recover this in the 2-dimensional case by taking
%------------------------
\[
f(r) = \begin{cases} 0 & r \geq r_0\\
\frac{1}{\pi r_0^2} & 0 \leq r < r_0
\end{cases}
\]
%------------------------
with
%------------------------
\[
\kappa = \rho \pi r_0^2
\]
%------------------------
%\cite{penrose2003random}, Chung--Lu random graphs~\cite{chung:connected}, Small world networks~\cite{watts:collective,newman1999renormalization}, Long-range percolation~\cite{X}, and Waxman graph~\cite{X}
for all nodes.  If the distance between $u$ and $v$ is less than $r_0$, then an edge exists with probability
%------------------------
\[
\frac{\kappa_u\kappa_v f(d_{uv})}{\rho \ave{\kappa}} = (\rho \pi r_0^2)^2 \frac{1}{(\pi r_0^2 \rho) (\rho \pi r_0^2)}  = 1 \ ,
\]
%------------------------
while if their distance is at least $r_0$, then no edge exists.  It is straightforward to generalize this to higher dimension.

%------------------------
\subsubsection{Chung--Lu and \erdosrenyi{} Graphs}
%------------------------
In a Chung--Lu network, an edge between $u$ and $v$ exists with probability $\kappa_u\kappa_v/(N\ave{\kappa})$ \cite{chung:connected}.  By setting $f(d) = 1/|V|$ (that is indepedent of $d$) we loose spatial structure.  Then an edge exists between $u$ and $v$ with probability 
%------------------------
\[
\frac{\kappa_u\kappa_v f(d_{uv})}{\rho \ave{\kappa}} = \kappa_u \kappa_v \frac{1}{\rho |V| \ave{\kappa}} = \kappa_u \kappa_v \frac{1}{N \ave{\kappa}} \ .
\]
%------------------------
Thus we reproduce the Chung--Lu networks.  If we further set $\kappa_u=\kappa$ to be fixed for all $u$, we arrive at the \erdosrenyi{} graph model introduced by Gilbert~\cite{gilbert1959random}.

\subsubsection{Lattice-based models}
%------------------------
We can finally consider network classes for which the nodes are placed at regular intervals.  To match these we must modify the RSNs to place nodes at lattice points.  We describe Long-Range Percolation here; and the Newman--Watts network class was described in the main text.

If nodes are placed at lattice points of $\Re^n$ and $f$ is taken to be $f(r) = 1/r^s$ for some exponent $s$, then this produces the ``long-range percolation'' model on lattices~\cite{coppersmith2002diameter}.

%------------------------
\subsection{Fast Network Generation}
%------------------------
At first sight, generating networks from the RSN class appears to be an $\order(N^2)$ operation as each of the $\binom{N}{2}$ edges exists independently of the others.  However, by modifying methods developed to generate \erdosrenyi{} and Chung-Lu  networks~\cite{miller:chunglu,batagelj:efficient} in linear time, we arrive at a much more efficient process. This makes large RSNs practical for simulation and analysis.

We briefly outline the method, under the assumption that the distance kernel is bounded above and is monotonically decreasing.  We divide the space $V$ into regions and order the nodes in each region by decreasing expected degree.  

We consider a region $X_i$, and define $u$ to be the first node in that region.  The probability that $u$ will have an edge with any subsequent node is bounded above by $p_0=\min(1,\kappa_u^2 f_{\text{max}}/\ave{\kappa})$ where $f_{\text{max}}= f(0)$ is the maximum of $f$. We then choose a number $r$ from a geometric distribution with probability $p_0$.  

We set $v_1$ to be the $r$th node following $u$ in $X_i$.  This is equivalent to $v_1$ being the first node chosen when sequentially considering each node with probability $p_0$.  Thus the ``candidate neighbor'' $v_1$ is chosen with probability $p_0$ which is at least $q=\min(1,\kappa_u \kappa_{v_1} f(d_{uv})/\ave{\kappa})$.  It is possible that $v_1$ is a ``false positive''.  To decide this, we generate a new random number uniformly from $(0,1)$, and if it is less than $q/p_0$, we decide that $v_1$ was correctly chosen, and we add an edge between them.  Otherwise we do not.

We then enter an iterative step.  After processing $v_i$, we set $p_i = \min (1,\kappa_u \kappa_{v_i} f_{\text{max}}/\ave{\kappa})$ and jump to the next candidate neighbor $v_{i+1}$.   Again $v_{i+1}$ may be a false positive, and we place an edge between $u$ and $v_{i+1}$ with probability $q/p_i$ where $q = \min (1,\kappa_u \kappa_{v_{i+1}} f(d_{uv_{i+1}})/\ave{\kappa})$.  As the iteration progresses, $\kappa_{v_i}$ decreases so $p_i$ decreases, and the jumps become larger.  The edges within each region are assigned in linear time.

We next place edges between the node $u$ and other regions $X_j$.  We find the minimum distance between $u$ and $X_j$ and use it to define $f_{\text{max}}$.  We take $w$ to be the first node in $X_j$.  We define $p_0 = \min \{ 1, \kappa_u \kappa_w f_{\text{max}}/\ave{\kappa}\}$.  We choose $v_1$ from $X_j$ as before, and iterate.  For $X_j$ farther from $u$ the jumps are larger.

%------------------------
%\section{Graph definition and properties}
%------------------------

%------------------------
\subsection{Governing equations for SIR disease}
%------------------------
%\
%\JCM{Joel, can you further streamline this in conjunction with section 4A?}

In this section we give a detailed derivation of the partial integro-differential equation (PIDE) that describes SIR disease propagation on RSNs, as introduced in the main text.

In many random network classes, it is possible to develop powerful mathematical approaches by taking advantage of the fact that as $N \to \infty$, the clustering of the network goes to zero.  This also occurs for our random spatial graphs.  If we hold all other parameters the same, but increase the node density $\rho$, the probability that any two neighbors $v$ and $w$ of node $u$ will themselves be neighbors scales like $1/\rho$. We use this to develop equations for SIR disease spread, following~\cite{volz:cts_time,miller:ebcm_overview,miller:initial_conditions}, but extended to the spatial setting.

We assume that the population-scale dynamics are deterministic.  That is, recognizing that the underlying process is stochastic, we assume that, as a function of $\vec{x}$, the proportion infected at some later time $t$ is uniquely determined from the initial proportion infected (as a function of $\vec{x}$).  This assumption becomes reasonable in the limit of a large network, and is effectively the continuum assumption.  We make the observation that the following four questions have identical answers, if indeed the population-scale dynamics are deterministic:
%------------------------
\begin{enumerate}
\item What fraction of nodes are susceptible, infected, or recovered at time $t$, as a function of location $\vec{x}$?
\item What is the probability a random node is susceptible, infected, or recovered at time $t$, as a function of location $\vec{x}$?
\item What is the probability a random node is susceptible, infected, or recovered at time $t$, as a function of location $\vec{x}$, given the system's state at time $0$?
\item What is the probability a randomly chosen test node $u$ is susceptible, infected, or recovered given the system's state at time $0$ if we prevent $u$ from transmitting to its neighbors?
\end{enumerate}
%------------------------
The first two questions are clearly equivalent.  The second and third are equivalent because we assume that the population-scale dynamics are deterministic.  The third and fourth are equivalent because as long as $u$ is susceptible, the fact it does not transmit is irrelevant, and once $u$ is infected, its recovery time is not affected by any transmissions it causes. So the requirement that $u$ does not transmit does not influence the evolution of the state of $u$; it only influences the states of neighbors of $u$.  The process we analyze when preventing $u$ from transmitting (question 4) is different from the process in questions 1--3, but the probability we obtain in answering question 4 is the same as the probability that is the answer to question 3. Note that the equivalence of questions 3 and 4 holds specifically for SIR disease; it is crucial for our argument that infected individuals cannot become susceptible again.
%, and it has a negligible effect overall in the continuum limit.

The final question is one that we can address using probabilistic tools.  We define a test node $u$ which is randomly selected in the network at location $\vec{x}$ and prevented from transmitting to its neighbors.  Under the assumption that $\rho$ is large, we can assume independence of $u$'s neighbors because clustering is negligible and because $u$ does not form a path of infection between its neighbors, precisely because we prevent it from transmitting.  
We seek to find the probability $u$ is susceptible, from which we will deduce the probability it is infected or recovered.
%\HDS{Joel, maybe you can explain me when we meet next, why precisely do we require that $u$ does not transmit? Is it to prevent the path of infection between neighbors, which would destroy independence? Where precisely do we use this in what follows?}

We pass to a continuum limit and define $S(\vec{x},t)$, $I(\vec{x},t)$ and $R(\vec{x},t)$ to be the probability that a test node $u$ placed at $\vec{x}$ would be susceptible, infected, or recovered at time $t$.  We similarly define  $\Phi_S(\vec{x},t)$, $\Phi_I(\vec{x},t)$, $\Phi_R(\vec{x},t)$, and $\Theta(\vec{x},t)$ such that: $\Phi_S$ is the probability a random neighbor of the test node $u$ is susceptible at time $t$, $\Phi_I$ is the probability the neighbor is infected but has not transmitted to $u$, $\Phi_R$ is the probability the neighbor has recovered without transmitting to $u$ and $\Theta=\Phi_S+\Phi_I+\Phi_R$ is, thus, the probability that a random neighbor has not transmitted to $u$.  

We assume infection is introduced to the population at time $t=0$ with $S(\vec{x},0)$ denoting the probability a node at $\vec{x}$ would be susceptible. 
%\HDS{Joel, do we also assume $R(\vec{x},0)=0$? It seems we need this below to compute $\Phi_R$, because $\Phi_R(\vec{x},0)=0$ is needed such that $1-\Theta(\vec{x},0)=\Phi_R(\vec{x},0)$, or is that not required?}
Because $k$, the number of neighbors $u$ has, is Poisson distributed with mean $\kappa_u$ (the probability of a given $k$ is $e^{-\kappa_u}\kappa_u^k/k!$), the probability $u$ is susceptible at a later time $t$ is $S(\vec{x},0) \sum_{k=0}^\infty e^{-\kappa_u}\kappa_u^k \Theta^k/k! = S(\vec{x},0) \exp[-\kappa_u(1-\Theta)]$.  
Note that we have used here that the neighbors are independent (since we prevent $u$ from transmitting).
If we do not take $\kappa_u$ as given, the probability $u$ is susceptible is
%------------------------
\[
S = S(\vec{x},0) \, \Psi(\Theta(\vec{x},t)) = S(\vec{x},0) \int_0^\infty  e^{-\kappa(1-\Theta(\vec{x},t))} P(\kappa) \, \mathrm{d}\kappa,
\]
%------------------------
with $\Psi(\Theta(\vec{x},t)) \equiv \int_0^\infty  e^{-\kappa(1-\Theta(\vec{x},t))} P(\kappa) \, \mathrm{d}\kappa$.
We augment this with $I = 1-S-R$ and $\dot{R} = \gamma I$.  We must still find $\Theta(\vec{x},t)$.

%------------------------
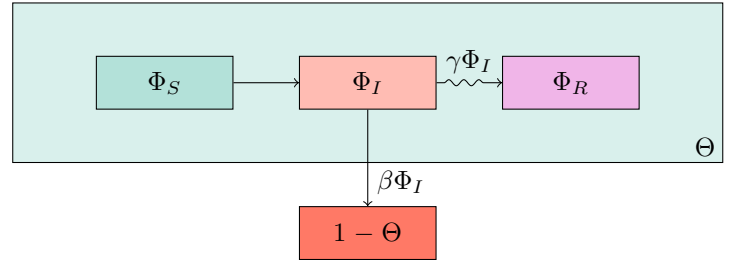
\begin{figure}[h]
\begin{center}
  \begin{tikzpicture}[xscale=0.9]
    \node[shortbox, fill=colorS!30] (phiS) at ( -2,1) {$\Phi_S$};%=\frac{\hat{\psi}'(\theta)}{\ave{K}}$};
    \node[shortbox, fill=colorI!30] (phiI) at ( 1,1) {$\Phi_I$};%=\theta-\phi_S-\phi_R$};
    \node[shortbox, fill=colorR!30] (phiR) at ( 4,1) {$\Phi_R$};
    \begin{scope}[on background layer]
      \node[bigbox = {$\Theta$}, fit=(phiS)(phiI)(phiR), fill=colorS!15] (Theta) {};
    \end{scope}
    \node[shortbox,fill=colorI!60] (OmTheta) at (1,-1) {$1-\Theta$};
    \path [->] (phiS) edge node {} (phiI);
    \path [decay] (phiI)--(phiR) node [midway, above] {$\gamma \Phi_I$};
    \path [->, right, near end] (phiI) edge node {$\beta \Phi_I$}
    (OmTheta);
  \end{tikzpicture}
\end{center}
\caption{\small A flow diagram leading to the governing equation for $\Theta(\vec{x},t)$.  The compartments $\Phi_S$, $\Phi_I$, and $\Phi_R$ make up $\Theta$ and represent the probability that a random partner has not transmitted to $u$ and is susceptible, infected, or recovered respectively.}
\label{fig:flow}
\end{figure}
%------------------------

We turn to the flow diagram in Fig.~\ref{fig:flow}.  Once a neighbor $v$ of the test node $u$ becomes infected, whether or not $v$ transmits infection to $u$ and if so, at what time the transmission occurs is independent of anything else that happens in the population.  Thus we can immediately calculate the flux from $\Phi_I$ to $1-\Theta$ is $\beta\Phi_I$, so $\dot{\Theta} = -\beta \Phi_I$.  Similarly we find the the flux to $\Phi_R$  is $\gamma \Phi_I$.  Since the flux to $1-\Theta$ and $\Phi_R$ are both proportional to $\Phi_I$ and both are $0$ at $t=0$, we can conclude that $\Phi_R = \gamma(1-\Theta)/\beta$.  However,  the rate at which $v$ becomes infected is more difficult.  It will be easier to directly calculate $\Phi_S$ in terms of $\Theta$ and use $\Phi_I = \Theta-\Phi_S-\Phi_R$ to give an equation for $\dot{\Theta}$ in terms of $\Theta$.

To find $\Phi_S$, we consider the possible neighbors of $u$ and calculate their probability of being susceptible.  The probability a node $v$ at $\hat{\vec{x}}$ is a neighbor of $u$ is proportional to $\kappa_v$ and to $f(|\hat{\vec{x}}-\vec{x}|)$.  In turn, the probability $v$ is susceptible is $S(\hat{\vec{x}},0) \exp[-\kappa_v(1-\Theta(\hat{\vec{x}},t))]$.  So the probability a random neighbor is susceptible is given by
%------------------------
\begin{align*}
\Phi_S &= \int_V S(\hat{\vec{x}},0)\int_\kappa \frac{\kappa P(\kappa)}{\ave{\kappa}} f(|\hat{\vec{x}}-\vec{x}|) e^{-\kappa(1-\Theta(\hat{\vec{x}},t))} \, \mathrm{d}\kappa \, \mathrm{d}\hat{\vec{x}}\\
&= \frac{\int_V S(\hat{\vec{x}},0) \Psi'(\Theta(\hat{\vec{x}},t)) f(|\hat{\vec{x}}-\vec{x}|) \, \mathrm{d}\hat{\vec{x}}}{\ave{\kappa}}.
\end{align*}
%------------------------
From this, $\dot{\Theta} = -\beta\Phi_I=-\beta(\Theta-\Phi_R-\Phi_S)$ becomes
%------------------------
\begin{align*}
\dot{\Theta}(\vec{x},t) &= - \beta \Theta(\vec{x},t) + \gamma(1-\Theta(\vec{x},t))\\
&\quad + \beta \frac{\int_{V}S(\hat{\vec{x}},0) \Psi'(\Theta(\hat{\vec{x}},t)) f(|\hat{\vec{x}}-\vec{x}|) \, \mathrm{d}\hat{\vec{x}}}{\ave{\kappa}}.
\end{align*}
%------------------------

\subsection{Relation to the Fisher--KPP Equation}
We note that, under approximations that are appropriate in the case of a localized spatial kernel with $\Theta$ varying slowly in space we can convert~\eqref{sys:PIDE} into the Fisher--KPP equation $u_t = ru(1-u/K) + D u_{xx}$.  We demonstrate this in the 1D case.

We start by assuming that $S(x,0)$ is approximately $1$.  We then assume that $f$ is sufficiently localized and $\Theta$ varies sufficiently slowly that we can expand $\Psi'(\Theta(x,t))$ as a Taylor Series in $x$ to third order. 
\begin{align*}
\int_{-\infty}^\infty \Psi'(\Theta(\hat{x},t)) &f(|\hat{x}-x|) \, \mathrm{d}\hat{x} \\
\approx & \int_{-\infty}^\infty \Psi'(\Theta(x,t)) \, \mathrm{d}\hat{x}\\
& + \int_{-\infty}^\infty (\hat{x}-x)\pd{}{x}\Psi'(\Theta(x,t)) f(|\hat{x}-x|)\, \mathrm{d}\hat{x} \\
&+ \int_{-\infty}^\infty \frac{(\hat{x}-x)^2}{2}\left(\pds{}{x} \Psi'(\Theta(x,t)\right) f(|\hat{x}-x|) \, \mathrm{d}\hat{x} \\
=&\Psi'(\Theta(x,t)) + C\pds{}{x}\Psi'(\Theta(x,t)) 
\end{align*}
where $C = \int_{-\infty}^\infty y^2f(|y|)/2 \, \mathrm{d}y$, and we have used the fact that $\int_{-\infty}^\infty y f(|y|) \, \mathrm{d}y=0$ by symmetry.

So taking $S(x,0)=1$, our equation for $\dot{\Theta}$ is
\[
\dot{\Theta} = - \beta \Theta + \gamma(1-\Theta) + \beta \frac{\Psi'(\Theta)}{\ave{\kappa}} + \frac{\beta C}{\ave{\kappa}}\pds{}{x} \Psi'(\Theta)
\]
Now setting $u = 1-\Theta$ and assuming $u$ is small, we use further Taylor expansions and the fact that $\Psi^{(n)}(1) = \ave{\kappa^n}$ to find
\begin{align*}
\dot{u} &= \beta (1-u) - \gamma u - \beta \frac{\Psi'(1-u)}{\ave{\kappa}} - \frac{\beta C}{\ave{\kappa}} \pds{}{x} \Psi'(1-u)\\
&\approx \beta (1-u) - \gamma u - \beta + \beta\frac{u\ave{\kappa^2}}{\ave{\kappa}} - \beta \frac{u^2 \ave{\kappa^3}}{2\ave{\kappa}} - \frac{\beta C}{\ave{\kappa}} \pds{}{x}\Psi'(1-u)\\
&\approx \left(\beta \frac{\ave{\kappa^2}}{\ave{\kappa}}-\beta-\gamma \right) u  - \beta \frac{\ave{\kappa^3}}{2\ave{\kappa}}u^2 - \frac{\beta C}{\ave{\kappa}} \pds{}{x} \left(\ave{\kappa} - u \ave{\kappa^2}\right)\\
&= r u(1-u/K) + D \pds{}{x} u
\end{align*}
for appropriately chosen $r$, $K$, and $D$.  So the Fisher--KPP equation arises out of an approximation of our PIDE on RSNs. In general these approximations are not accurate when, e.g., $\Theta$ is not close to $1$, the variation in $\Theta$ is fast enough that the Taylor Series expansions are poor approximations, or the spatial kernel is not localized.  However, this relation does suggest that behaviors found for the Fisher--KPP equation are likely to occur for disease spread in our RSNs as well.

%%%%%%%%%%%%%%%%%%%%%%%%%%%%%%%%%%%%%%%%%%%%%%
%%%%%%%%%%%%%%%%%%%%%%%%%%%%%%%%%%%%%%%%%%%%%%
%%%%%%%%%%%%%%%%%%%%%%%%%%%%%%%%%%%%%%%%%%%%%%
\end{document}